\documentclass[superscriptaddress,aps,prl,reprint,twocolumn,amsmath,amssymb]{revtex4-2}
\usepackage{graphicx}
\usepackage{amssymb}
\usepackage{slashed}
\usepackage{dcolumn}
\usepackage{amsmath}
\usepackage{bm}
\usepackage{colordvi}
\usepackage{algorithm}
\usepackage{algpseudocode}
\usepackage{multirow}
\usepackage{titlesec}
\usepackage{newtxtext,newtxmath}
\usepackage{dsfont}
\usepackage{xcolor}
\usepackage{mathbbol}

\usepackage{hyperref}
\hypersetup{
    colorlinks=true,
    linkcolor=blue,
    citecolor=blue,     
    urlcolor=blue,
}

\allowdisplaybreaks

\usepackage{mathrsfs}
\makeatletter

\newcommand{\Rmnum}[1]{\expandafter\@slowromancap\romannumeral #1@}
\makeatother

\begin{document}

\title{Altermagnetism-induced non-collinear superconducting diode effect and unidirectional superconducting transport}

\author{F. Yang}
\email{fzy5099@psu.edu}

\affiliation{Department of Materials Science and Engineering and Materials Research Institute, The Pennsylvania State University, University Park, PA 16802, USA}

\author{L. Q. Chen}
\email{lqc3@psu.edu}

\affiliation{Department of Materials Science and Engineering and Materials Research Institute, The Pennsylvania State University, University Park, PA 16802, USA}

\date{\today}

\begin{abstract}

  Current studies of non-reciprocal superconducting (SC) transport have centered on the forward–backward asymmetry of the critical current measured along a single axis. In most realizations, this diode effect is achieved via introducing ferromagnetism or applying an external magnetic field, which drives system into an effective Fulde–Ferrell (FF) state but often at the cost of severely suppressing the SC gap and thus compromising device robustness. Here we propose and theoretically demonstrate that coupling a conventional $s$-wave SC thin film to a $d$-wave altermagnet offers a more resilient alternative.  The momentum-dependent spin splitting inherent to altermagnets induces a non-collinear SC-diode effect in the BCS state, with the critical-current anisotropy exhibiting a fourfold ($C_4$) symmetry. Upon entering the FF state at large splitting, this anisotropy gradually evolves into a unidirectional ($C_1$) pattern.  Crucially, the FF pairing momentum locks to the discrete crystal axes, eliminating the rotational Goldstone mode and preserving a sizable SC gap without any abrupt or significant suppression. These combined features make the altermagnetic proximity an appealing platform to engineer symmetry-protected, energy-efficient and programmable  SC diodes for next-generation  electronic devices.

\end{abstract}

\maketitle  

\begin{figure}[htb]
  {\includegraphics[width=6.0cm]{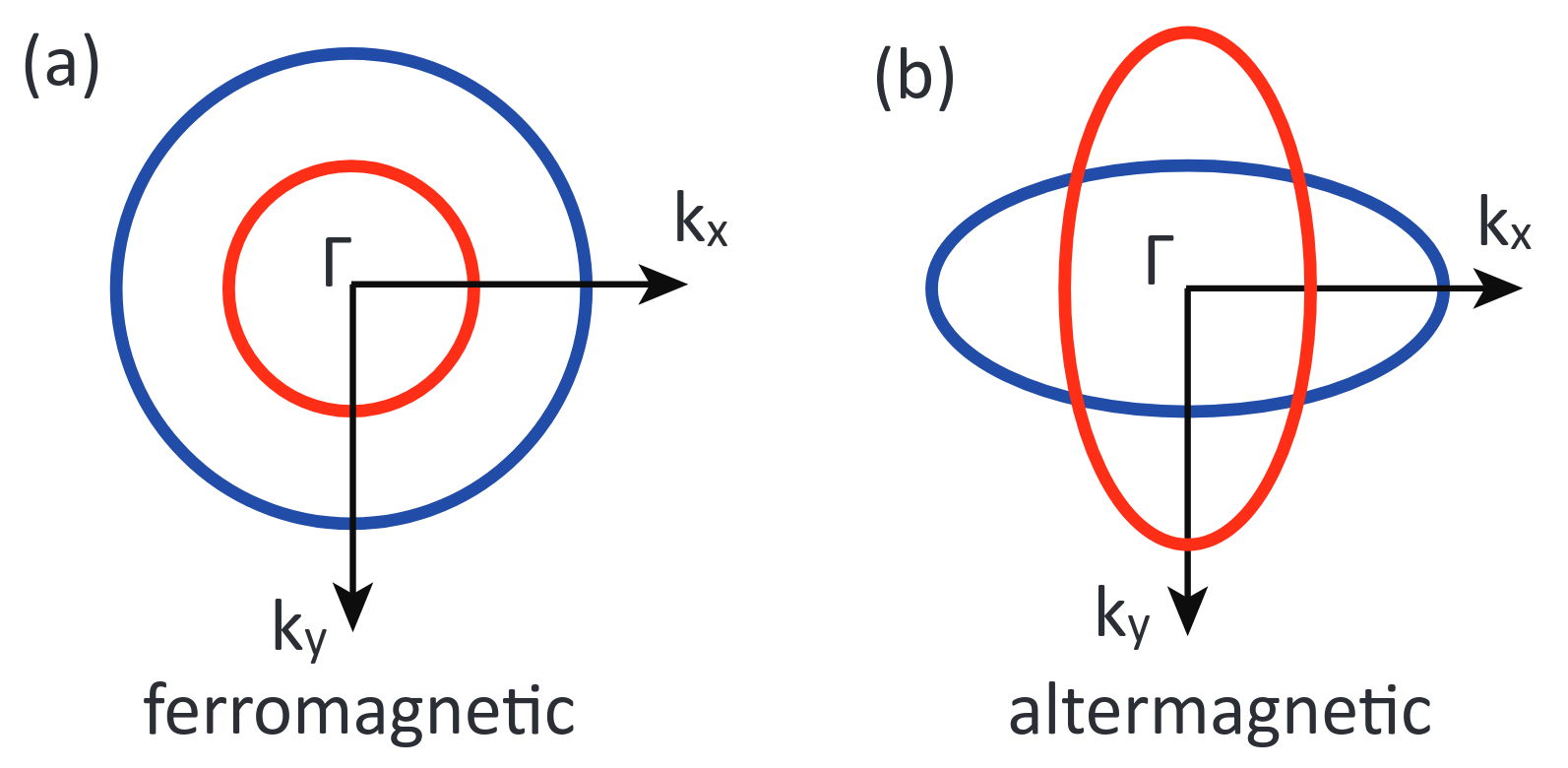}}
  \caption{Schematic of spin-up and spin-down Fermi surfaces under ferromagnetic (left) and $d$-wave altermagnetic (right) spin splitting.} 
\label{figyc1}
\end{figure}

{\sl Introduction.---}Non-reciprocal superconducting (SC) transport, in which the supercurrent flows without dissipation in one direction but encounters finite resistance in the reverse, has attracted considerable and growing interest~\cite{PhysRevX.12.041013,PhysRevLett.128.037001,PhysRevLett.128.177001,doi:10.1073/pnas.2119548119,He_2022,Scammell_2022,doi:10.1126/sciadv.abo0309,PhysRevB.106.214524,bauriedl2022supercurrent,ando2020observation,wu2022field,PhysRevLett.121.026601,PhysRevB.106.205206,nagaosa2024nonreciprocal} for its potential applications in next-generation quantum and ultra-low-dissipation electronic technologies. This phenomenon, known as the SC diode effect, enables directional control of supercurrents and offers a promising pathway toward inherently programmable SC circuits.  The underlying microscopic mechanism so far is associated with the  finite-momentum Cooper pairing~\cite{doi:10.1073/pnas.2119548119}.  Realizing such a state typically requires breaking time-reversal symmetry, for instance, by introducing ferromagnetism or applying an external magnetic field. The resulting Zeeman spin splitting leads to mismatched Fermi surfaces for spin-up and spin-down electrons [Fig.~\ref{figyc1}(a)], favoring SC pairs with a finite center-of-mass (CM) momentum and thus driving a Fulde–Ferrell–Larkin–Ovchinnikov (FFLO) transition~\cite{larkin1965zh,fulde1964superconductivity}.  As a consequence, the critical current for superconductivity along the pairing CM momentum differs from that in the opposite direction, giving rise to the diode effect. 

However, several challenges hinder the practical implementation of this mechanism.  The external magnetic fields typically employed to induce Zeeman splitting often introduce the additional orbital effect, and hence suppress superconductivity by generating vortices or pair-breaking currents~\cite{abrikosov2012methods}. On the other hand, in the original FFLO theory, the direction of the Cooper-pair CM momentum is energetically degenerate~\cite{fulde1964superconductivity}, such that the emergence of a CM momentum spontaneously breaks the continuous rotational symmetry and gives rise to a gapless rotational Goldstone mode~\cite{goldstone1961field,goldstone1962broken}. This, in principle, renders the conventional FFLO state inherently fragile and difficult to stabilize. To address this issue, various theoretical proposals have introduced spin–orbit coupling (SOC) to lift the degeneracy and pin the CM momentum direction~\cite{yang2018fulde,Dong_2013,PhysRevA.89.013607,PhysRevLett.89.227002,PhysRevB.76.014522,PhysRevB.75.064511,PhysRevLett.114.110401}. However, the first-order FFLO transition is usually accompanied by a substantial SC gap suppression~\cite{yang2018fulde,fulde1964superconductivity,Dong_2013,PhysRevB.111.054501}, weakening the condensate and undermining the robustness required for diode functionality.  This intrinsic trade-off severely limits the practical viability of the FFLO-based SC diodes.

Very recently, a novel class of magnetic materials, known as altermagnets, has been theoretically proposed and possibly observed experimentally~\cite{vsmejkal2022beyond,vsmejkal2022emerging,bhowal2024ferroically,reichlova2024observation,ding2024large,PhysRevLett.133.156702,zhang2025crystal,lin2024observation}. Altermagnets feature oppositely spin-polarized sublattices related by crystalline rotation symmetries. While they break time-reversal symmetry like ferromagnets, they exhibit no net magnetization due to the sublattice compensation, similar to antiferromagnets~\cite{vsmejkal2022beyond,vsmejkal2022emerging,bhowal2024ferroically}. 
 A key characteristic of altermagnetism is an even-parity,  momentum-dependent spin splitting that respects inversion symmetry and originates from symmetry-protected band degeneracies,  typically exhibiting $d$-, $g$-, or $i$-wave form~\cite{vsmejkal2022beyond,vsmejkal2022emerging}, resulting in   unconventional spin textures, as illustrated in Fig.~\ref{figyc1}(b) for a $d$-wave case. These distinct features position altermagnets as highly promising platforms for both spintronic~\cite{bai2024altermagnetism,weissenhofer2024atomistic,gonzalez2021efficient,PhysRevLett.130.216701,sun2025spin} and quantum transport applications~\cite{feng2022anomalous,liao2024separation,sato2024altermagnetic,ma2021multifunctional,fu2025all}.

Given their time-reversal-breaking yet inversion-symmetric character, and the absence of the orbital effects due to a net  magnetization, altermagnets provide a natural platform for exploring their interplay with superconductivity~\cite{fukaya2025superconducting,fukaya2025josephson,PhysRevB.111.184515,PhysRevLett.131.076003,PhysRevB.108.054511,dlpb-gfct}. The intrinsic momentum-dependent spin polarization in altermagnets induces a mismatch between spin-up and spin-down Fermi surfaces, favoring the finite-momentum Cooper pairing even in the absence of external fields~\cite{zhang2024finite,hu2025unconventional}. Recent theoretical works~\cite{6wxh-p4mc,mukasa2025finite,sim2024pair,PhysRevResearch.5.043171,hu2025quantum} have investigated the stability of such pairing near the SC transition, typically by solving the linearized gap equation. A fully self-consistent microscopic analysis was performed by Hong~\textit{et al.}~\cite{PhysRevB.111.054501}, who systematically compared conventional BCS pairing, chiral $p$-wave states, and FF pairing under comparable altermagnetic conditions. Their results revealed that, within a narrow but realistic range of altermagnetic strength, the FF state can become energetically favorable without requiring external magnetic fields or fine-tuned SOC.

Motivated by these developments~\cite{hu2025unconventional,zhang2024finite,PhysRevB.111.054501}, here we investigate directional SC transport in conventional $s$-wave SC thin films proximity-coupled~\cite{PhysRevB.95.075304,PhysRevB.84.144522,PhysRevB.91.165425,PhysRevLett.118.117001,RevModPhys.77.935} to a $d$-wave altermagnet, using a fully microscopic and self-consistent framework.  As the altermagnetic strength increases from zero,   we find that the critical current in the BCS regime evolves from being isotropic to strongly anisotropic with a fourfold ($C_4$) symmetry, reflecting the underlying crystalline symmetry of the altermagnetic spin texture. Thus, the directions of maximal and minimal critical currents are not antiparallel, indicating that the non-reciprocity deviates from the simple single-axis diode behavior and instead reflects   a {\sl non-collinear} SC diode effect. 

Upon further increase of the altermagnetic strength, the system undergoes a transition into the FF state, where Cooper pairs acquire a finite CM momentum. In this regime, the critical-current contour gradually develops into a pronounced unidirectional ($C_1$) form aligned with the CM momentum, i.e., a single SC transport direction is spontaneously selected and pinned. Notably, unlike the conventional FFLO mechanism~\cite{yang2018fulde,fulde1964superconductivity,Dong_2013}, where the emerging CM momentum from the Zeeman-split Fermi surfaces spontaneously breaks the continuous rotational symmetry, the momentum-dependent spin splitting in altermagnets is intrinsically encoded in the crystal lattice. As a result, the optimal pairing momentum is locked to a discrete set of directions (four in the $d$-wave case), eliminating the associated rotational Goldstone mode~\cite{goldstone1961field,goldstone1962broken} and stabilizing a directionally pinned SC state. Importantly, the BCS-to-FF transition driven by the altermagnetic splitting does not exhibit the abrupt and strong gap suppression typically observed in the conventional FFLO phases~\cite{yang2018fulde,fulde1964superconductivity,Dong_2013}.  These features make altermagnets a promising platform for realizing robust, directionally controllable SC diode effects while maintaining a sizable SC gap and strong superfluid density, ensuring the operational stability and reliable integration required for high-performance, scalable SC electronics.

{\sl Model.---}We begin with a conventional $s$-wave SC Hamiltonian, incorporating the $d$-wave altermagnetic spin splitting induced via proximity coupling~\cite{PhysRevB.95.075304,PhysRevB.84.144522,PhysRevB.91.165425} to a $d$-wave altermagnet (such as RuO$_2$\cite{lin2024observation}, ${\mathrm{Mn}}_{5}$${\mathrm{Si}}_{3}$~\cite{reichlova2024observation} and $\beta$-MnO$_2$~\cite{vsmejkal2022emerging,noda2016momentum}). The Hamiltonian is written as~\cite{abrikosov2012methods,schrieffer1964theory,bardeen1957theory,PhysRevB.95.075304} 
\begin{equation}
H=\!\sum_{{\bf k}s}(\xi_{\bf k}+sh_{\bf k}){\hat n}_{{\bf k}s}+\!\sum_{\bf kk'k_1k_1'}U_{\bf kk'k_1k_1'}c_{{\bf k}\uparrow}^{\dagger}c_{{\bf k'}\downarrow}^{\dagger}c_{{\bf k_1}\downarrow}c_{{\bf k_1'}\uparrow},  
\end{equation}
where number operator ${\hat n}_{{\bf k}s}=c_{{\bf k}s}^{\dagger}c_{{\bf k}s}$ with $c_{{\bf k}s}^{\dagger}$ and $c_{{\bf k}s}$ being the electron creation and annihilation operators for momentum ${\bf k}$ and spin   $s=\pm$ ($\uparrow$ and $\downarrow$); $\xi_{\bf k}={\hbar^2k^2}/({2m})-\mu$ is the kinetic energy measured from the chemical potential $\mu$, with $m$ being the effective mass of electrons; $h_{\bf k}=\gamma_d(k_x^2-k_y^2)$ denotes the $d$-wave momentum-dependent spin splitting~\cite{sun2025spin,vsmejkal2022beyond,vsmejkal2022emerging,lin2024observation,PhysRevB.111.184515}  arising from the $d$-wave altermagnetic proximity effect. The second term describes a generalized spin-conserving interaction, which, for an on-site $s$-wave pairing interaction~\cite{li21superconductor}, takes the form $U_{\bf kk'k_1k_1'}=U\delta({\bf k+k'- k_1-k_1'})$.

We consider the phase-modulated FF state~\cite{yang2018fulde,fulde1964superconductivity,Dong_2013,PhysRevB.111.054501,yang24thermodynamic}, characterized by a SC order parameter with a single finite CM momentum ${\bf q}$, i.e., all Cooper pairs formed between two electrons with opposite spins and momenta ${\bf k}$ and ${\bf k}'$ condense at the same finite CM momentum ${\bf q}$, in order to maximize the available pairing phase space. Unlike pair-density-wave states (such as the multiple-${\bf q}$ LO state~\cite{larkin1965zh}), which exhibit a spatial modulation of the gap and break translational symmetry, the FF state preserves the translational invariance of the SC gap, featuring only a spatially varying SC phase. The corresponding SC gap is given by $\Delta_{\bf q}=U\sum_{\bf k}\langle{c_{{\bf -k+q}\downarrow}c_{{\bf k+q}\uparrow}}\rangle$, which leads to the mean-field Hamiltonian: 
\begin{eqnarray}
  H_{\bf q}&=&\sum_{\bf k}\big[(\xi_{\bf k+q}+h_{\bf k+q}){\hat n}_{{\bf k+q}\uparrow}+(\xi_{\bf -k+q}-h_{\bf -k+q}){\hat n}_{{\bf -k+q}\downarrow}\nonumber\\
  &&+\Delta_{\bf q}c_{{\bf -k+q}\downarrow}c_{{\bf k+q}\uparrow}+\Delta_{\bf q}c_{{\bf k+q}\uparrow}^{\dagger}c_{{\bf -k+q}\downarrow}^{\dagger}\big]-\frac{\Delta_{\bf q}^2}{U}.
\end{eqnarray}
{Diagonalizing this Hamiltonian via Bogoliubov transformation, one obtains 
$H_{\bf q}=\sum_{\bf k}(E^+_{\bf k}\alpha_{\bf k}^{\dagger}\alpha_{\bf k}-E_{\bf k}^-\beta_{\bf k}^{\dagger}\beta_{\bf k}-E_{\bf k}2V_{\bf k}^2)-\frac{\Delta_{\bf q}^2}{U}$.  
The SC free energy of the system is given by~\cite{PhysRevB.58.9365} 
\begin{eqnarray}
\mathcal{F}^s_{\mathbf{q}}&=&\langle{H}\rangle+k_BT\sum_{{\bf k},\lambda=\pm,\eta=\pm}\big[f(\eta\lambda{E_{\bf k}^{\lambda}})\ln{f(\eta\lambda{E_{\bf k}^{\lambda}})}\big]\nonumber\\
 &=&\sum_{{\bf k},\lambda=\pm}\big[(E_{\bf k}+\lambda\Gamma_{\bf k})f(E_{\bf k}+\lambda\Gamma_{\bf k})+k_BT{\ln}f(E_{\bf k}+\lambda\Gamma_{\bf k})\nonumber\\
 &&+(E_{\bf k}+\lambda\Gamma_{\bf k})\big(1-f(E_{\bf k}+\lambda\Gamma_{\bf k})\big)\big]-\sum_{\bf k}E_{\bf k}2v_{\bf k}^2-\frac{\Delta_{\bf q}^2}{U}\nonumber\\
 &=&-\!\frac{1}{\beta}\sum_{\mathbf{k},\lambda = \pm} \ln\big(1 \!+\! e^{-\beta\lambda E_{\mathbf{k}}^\lambda} \big) \!-\! \sum_{\mathbf{k}} \left(E_{\bf k}\!-\! \Xi_{\mathbf{k}}\right) \!-\! \frac{|\Delta_{\mathbf{q}}|^2}{U},
\end{eqnarray}
where $\beta = 1/(k_B T)$. 
Minimizing $\mathcal{F}^s_{\bf q}$ with respect to $\Delta_{\bf q}$ yields the self-consistent gap equation:
\begin{equation}\label{GE}
\frac{1}{U}=\sum_{\bf k}\frac{f(E_{\bf k}^{+})-f(E^-_{\bf k})}{2E_{\bf k}},  
\end{equation}
where the quasiparticle energy spectrum is given by $E_{\bf k}^{\pm}=\hbar{\bf v}_{\bf k}\cdot{\bf q}+h_{\bf k}\pm{E_{\bf k}}$ with ${\bf v}_{\bf k}=\hbar{\bf k}/m$ and $E_{\bf k}=\sqrt{\Xi_k^2+\Delta_{\bf q}^2}$~\cite{PhysRevB.111.054501,hu2025unconventional}. Here, $\Xi_{\bf k}=\xi_{\bf k}+2\gamma_d(k_xq_x-k_yq_y)$, suggesting a moving-frame    transformation~\cite{PhysRevResearch.6.033006}, while $f(x)$ is the Fermi-Dirac distribution function, and the condensate density $2{V_{\bf k}}^2=1-\Xi_{\bf k}/E_{\bf k}$.}  The quasiparticle energies here are tilted by the finite CM momentum. As a result, at $T=0$, the integrand in the gap equation becomes nonzero only when $E_{\bf k}^{+} > 0$ and $E_{\bf k}^{-} < 0$. Thus, SC anomalous pairing correlations vanish in momentum-space regions where either $E_{\bf k}^{+} < 0$ or $E_{\bf k}^{-} > 0$, as $f(E_{\mathbf{k}}^{+}) - f(E_{\mathbf{k}}^{-}) \equiv 0$. These regions are referred to as unpairing regions~\cite{fulde1964superconductivity,yang2018fulde,yang21theory}, where electrons no longer contribute to Cooper pairing and instead behave as normal-state excitations. 

 The equilibrium SC state is determined by self-consistently solving the gap equation and minimizing the free energy $\mathcal{F}^s_{\bf q}$ with respect to the pairing momentum ${\bf q}$. This procedure allows us to determine the optimal CM momentum ${\bf q}_0$ and the corresponding practical SC gap $\Delta=\Delta_{\bf q=q_0}$ as a function of system parameters such as the altermagnetic strength and temperature. 
 {To investigate the transport properties, we introduce an additional SC momentum 
 ${\bf p}_s=\nabla_{\bf R}\theta/2-e{\bf A}$~\cite{yang2018gauge,littlewood1981gauge,yang19gauge,nambu1960quasi,yang21theory,sun20collective,schrieffer1964theory,abrikosov2012methods,littlewood81gauge,ambegaokar61electromagnetic}, corresponding to a uniform phase gradient, superflow, or external vector potential. The resulting supercurrent is given by ${\bf j}_s = (e/m)n_s {\bf p}_s$~\cite{yang2018gauge,littlewood1981gauge,yang19gauge,nambu1960quasi,yang21theory,sun20collective,schrieffer1964theory,PhysRevB.102.144508}, where $n_s$ is the superfluid density~\cite{SDC}, which approaches the total electron  density $n$ in the low-temperature limit~\cite{sun20collective,yang24thermodynamic,yang2018fulde,yang2018gauge,abrikosov2012methods,PhysRevB.109.064508,PhysRevB.106.144509}. 
 To determine the critical current, we gradually increase the magnitude of ${\bf p}_s$ along a fixed direction and solve the gap equation self-consistently under the total momentum ${\bf q}_0 + {\bf p}_s$. The SC gap collapses at a critical value ${\bf p}_{s,c}$, signaling the pair-breaking threshold and marking the onset of dissipation.}

 For comparison, we also consider a conventional FFLO model with an isotropic spin splitting arising from ferromagnetic proximity, modeled as $h_{\bf k} = \gamma_s k^2$. This serves to isolate the role of the anisotropic, crystal-symmetry-protected spin splitting characteristic of altermagnets.

\begin{figure}[htb]
  {\includegraphics[width=8.7cm]{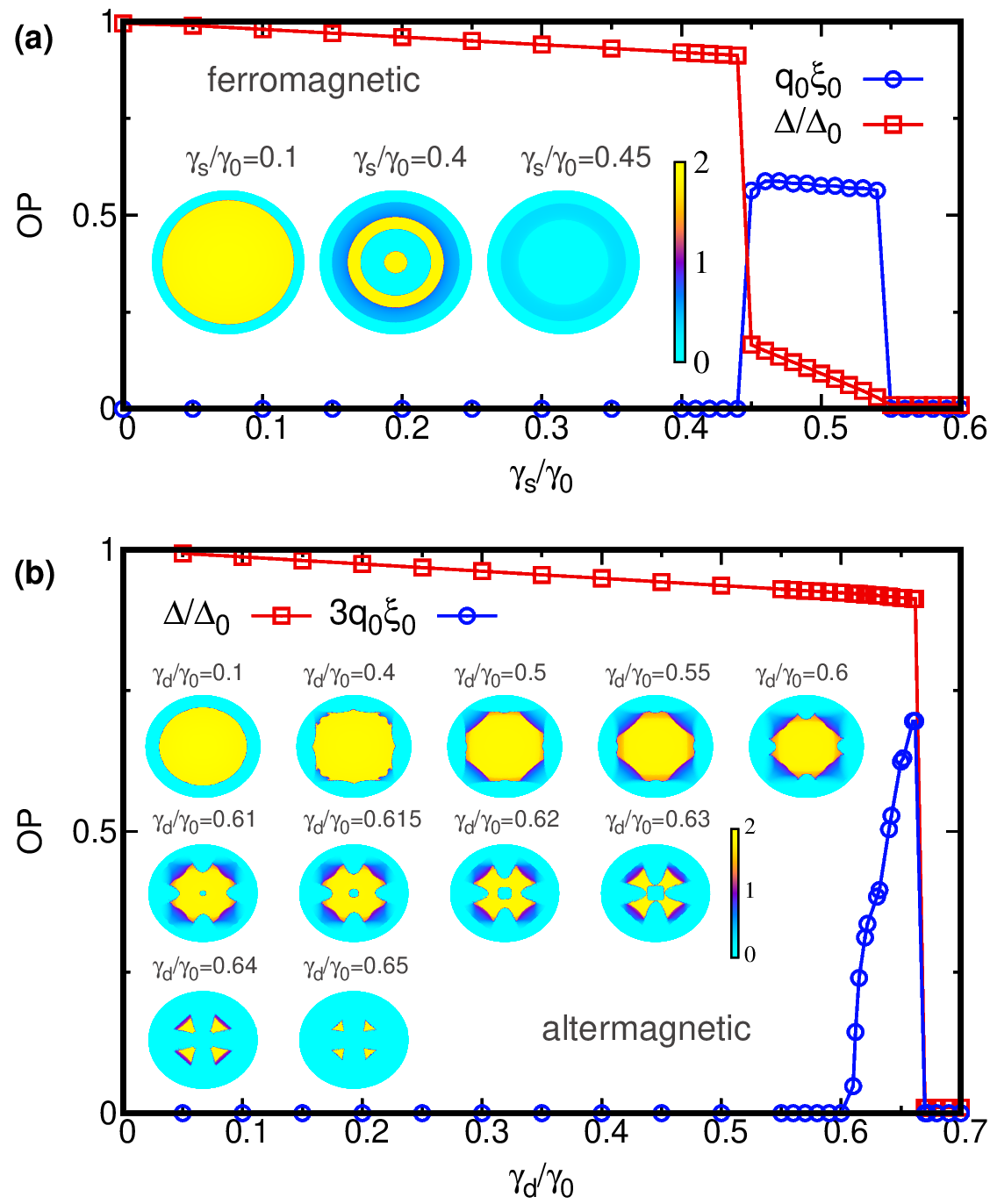}}
  \caption{Simulated order-parameter (OP) results for the optimal CM momentum $q_0$ and the corresponding practical SC gap $\Delta = \Delta_{\bf q=q_0}$ as functions of the spin-splitting strength for {\bf (a)} ferromagnetic and {\bf (b)} $d$-wave altermagnetic proximity. Insets: the $\Delta_{\bf q}$ landscape plotted in $q_x$–$q_y$ plane,  which reflects the inverse of the free-energy profile. The plotted $q$ range extends up to $0.8q_0$ in each direction. $T=0.1~$K. Normalized parameters used here are: the zero-splitting gap $\Delta_0=2~$meV, coherence length $\xi_0 = v_F/\Delta_0$, and splitting unit $\gamma_0 = \Delta_0/k_F^2$.} 
\label{figyc2}
\end{figure}

{\it Phase diagram.---} As a specific example, we consider a typical SC thin film, NbN, which has a SC gap $\Delta_0 \sim 2~\text{meV}$ at $T\sim0$ and a critical SC temperature $T_c\sim12$-15~K~\cite{yang24thermodynamic,chockalingam2008superconducting,matsunaga2014light}. Due to the Cooper instability of interacting fermions underlying the BCS theory~\cite{abrikosov2012methods,bardeen1957theory,schrieffer1964theory,mahan2013many}, the momentum summation for the pairing electrons is restricted such that the spin-up electron with momentum ${\bf k+q}$ and the spin-down electron with momentum ${\bf -k+q} $ both lie within their respective Debye momentum shells around the Fermi surface, enforced by the spin-dependent condition $|\xi_{s{\bf k+q}}+sh_{s{\bf k+q}}|\le\omega_D$. The CM momentum dependence of $\Delta_{\bf q}$ directly reflects the inverse of the free-energy landscape since $\mathcal{F}^s_{\mathbf{q}} - \mathcal{F}^n_{\mathbf{q}} \propto -\Delta_{\bf q}^2$, offering an efficient means for following analysis. 

The results for the ferromagnetic-proximity  case are shown in Fig.~\ref{figyc2}(a). As the Zeeman splitting $\gamma_s$ is raised from zero, $\Delta=\Delta_{\bf q=0}$ in the BCS state is only mildly suppressed due to the gradual reduction of the available pairing phase space. With further increase, as shown in the insets of Fig.~\ref{figyc2}(a), the $\Delta_{\bf q}$ landscape (the region with nonzero $\Delta_{\bf q}$) evolves from a single maximum at $q = 0$ (BCS state), to a double-well structure with local maxima at both $q= 0$ and $q\ne 0$, and eventually to a single maxima at a finite $q\ne 0$, signaling the formation of the FF state for $\gamma_s\ge0.45\gamma_0$. This evolution reflects a first-order phase transition~\cite{Landaubook}, and is accompanied by an abrupt and significant suppression of the SC gap $\Delta=\Delta_{\bf q=q_0}$. Owing to the energy degeneracy in the direction of ${\bf q}_0$, the FF state with an emerging CM momentum spontaneously breaks the continuous rotational symmetry and supports a gapless Goldstone mode~\cite{goldstone1962broken,goldstone1961field}, rendering it intrinsically fragile to fluctuations and disorder. At higher splitting, the remaining $\Delta_{\bf q}$ landscape continuously shrinks, and the system undergoes a second-order transition to the normal state at $\gamma_s = 0.55\gamma_0$. These results agree fully with the conventional FFLO theory~\cite{yang2018fulde,Dong_2013,fulde1964superconductivity}.

The altermagnetic-proximity case exhibits qualitatively distinct behaviors, as shown in Fig.~\ref{figyc2}(b). As the $d$-wave spin splitting $\gamma_d$ increases from zero, the SC gap $\Delta=\Delta_{\bf q=0}$ in the BCS state is only mildly suppressed. After a critical strength $\gamma_d > 0.6\gamma_0$, a finite CM momentum $q_0$ develops gradually, signaling a continuous transition to the FF state. Remarkably, this transition is not accompanied by any abrupt or substantial suppression of the SC gap $\Delta=\Delta_{\bf q=q_0}$. Both the gap reduction and the growth of $q_0$ evolve smoothly throughout the entire FF phase, in sharp contrast to the isotropic ferromagnetic case [Fig.~\ref{figyc2}(a)]. Beyond a second threshold $\gamma_d = 0.66\gamma_0$, the system undergoes a first-order transition directly from the FF state to the normal state, evidenced by a sudden collapse of the SC gap. This markedly different topology (a second-order transition to the FF state and a first-order transition to the normal state) totally arises from the strongly anisotropic $d$-wave spin splitting, which reconstructs the available pairing phase space in a fundamentally different way from the isotropic case.

As seen from the insets of Fig.~\ref{figyc2}(b), tracking $\Delta_{\bf q}$ landscape [the regions with nonzero $\Delta_{\bf q}$ including both yellow and purple ones]  as a function of increasing $\gamma_d$, we observe the following evolution. At zero splitting, the gap landscape is perfectly isotropic and circular, with a global maximum at ${\bf q}=0$, reflecting the full rotational symmetry of the BCS state. As $\gamma_d$ increases but remains below the critical value $0.6\gamma_0$, the circular contour gradually deforms into a nearly $C_4$-symmetric  square-like shape in ${\bf q}$-space, but with pronounced inward indentations along the spin-splitting axes (i.e., the $x$ and $y$ directions). This distortion mirrors the fourfold crystalline anisotropy of the $d$-wave altermagnet, although the global maximum remains pinned at the origin ($q=0$). Beyond the critical strength $0.6\gamma_0$, the gap at ${\bf q}=0$ (i.e., the BCS state) becomes fully suppressed, forming an isotropic depression around the origin. The system is thus poised to relocate the gap maximum to a finite momentum and enter the FF state. Immediately beyond this point, i.e., in the early FF regime, the $\Delta_{\bf q}$ landscape spans a continuous connected region with a circular depression at the center and pronounced inward indentations along the spin-splitting axes at the outer boundary. Thus, although the global maximum of $\Delta_{\bf q}$ has already shifted off center, the landscape still retains partial connectivity. As $\gamma_d$ increases further, this region begins to weaken in the continuity, and four incipient lobes become visible along the nodal directions of the splitting, in order to achieve available pairing phase space for the   $d$-wave-split Fermi surfaces in Fig.~\ref{figyc1}(b).   For $\gamma_d > 0.63\gamma_0$, the landscape fractures into four discrete maxima, each pinned to a nodal direction of the $d$-wave spin splitting, 
thereby locking the FF pairing momentum~\cite{PhysRevB.111.054501} and enhancing the robustness against fluctuations by eliminating rotational Goldstone modes~\cite{goldstone1962broken,goldstone1961field}. Eventually, once the spin-split Fermi surfaces cease to overlap, the four lobes contract and $\Delta_{\bf q}$ collapses, signaling a first-order transition to the normal state.

\begin{figure}[htb]
  {\includegraphics[width=8.6cm]{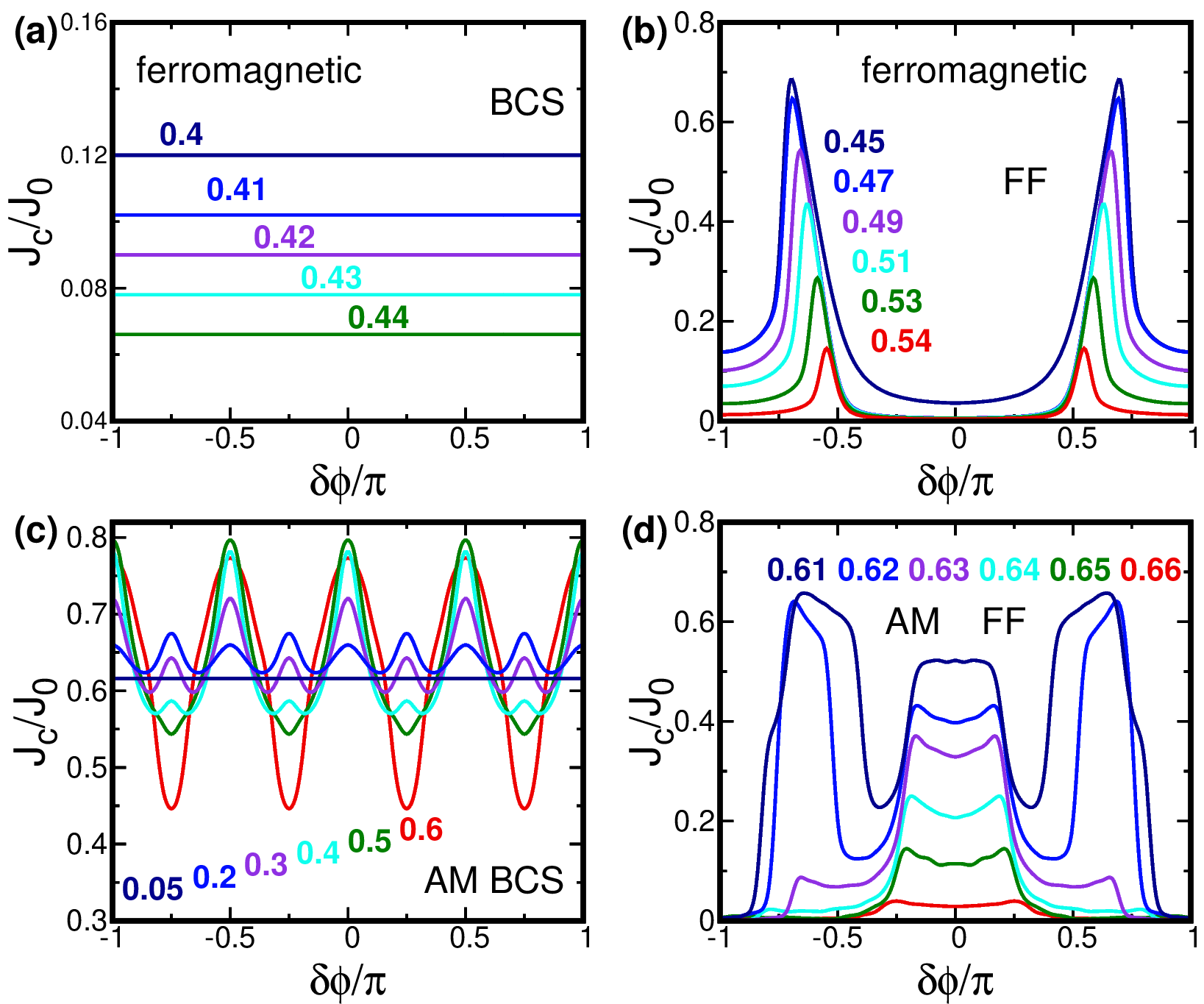}}
  \caption{De-pairing critical current $J_{c}(\delta\phi)$ as a function of the current direction at various spin splittings. (a) and (b): results for ferromagnetic proximity at different strengths  $\gamma_{s}/\gamma_0$ in BCS and FF regimes, respectively; (c) and (d): results for $d$-wave altermagnetic proximity at different strengths  $\gamma_{d}/\gamma_0$ in BCS and FF regimes, respectively. The characteristic current scale $J_0=en_s/(m\xi_0)$ and then $J_c/J_0=p_{s,c}\xi_0$. The angle $\delta\phi=\phi_{\bf p_s}-\pi/4$ and we set $\phi_{\bf q_0}=\pi/4$ throughout. } 
\label{figyc3}
\end{figure}

{\it Critical currents.---}After determining the phase diagram, we now present the key transport result of this study: the angular dependence of the de-pairing current $J_{c}(\delta\phi)$, plotted in Fig.~\ref{figyc3}.  The $\delta\phi = \phi_{\bf p_s} -\pi/4$ denotes the angle between the applied supercurrent direction and $\pi/4$ axis, which corresponds to one of the nodal directions of the $d$-wave altermagnetic spin splitting. As mentioned above, after the FF transition, the emerging optimal CM momentum ${\bf q}_0$ in the ferromagnetic-proximity case is directionally degenerate, whereas in the altermagnetic-proximity case it becomes locked to one of the discrete crystalline directions at $(2n+1)\pi/4$ axes. For concreteness, we set $\phi_{\bf q_0} =\pi/4$ throughout.

In the BCS phase at the ferromagnetic-proximity case  [Fig.~\ref{figyc3}(a)],  $J_c(\delta\phi)$ is fully isotropic and decreases gradually with increasing $\gamma_s$, as expected. Upon entering the FF phase via a first-order transition [Fig.~\ref{figyc3}(b)], the critical-current contour becomes strongly anisotropic, and a clear forward–backward asymmetry emerges on the ${\bf q}_0$ axis, i.e., $J_c(\delta\phi=0){\ne}J_c(\delta\phi=\pm\pi)$, manifesting as a SC-diode effect in agreement with earlier prediction~\cite{doi:10.1126/sciadv.abo0309}. Interestingly, the largest critical currents do not occur precisely at $\delta\phi = \pi$, but instead appear in the range $\delta\phi \approx 0.75$–$0.5\pi$, while the minimum lies at $\delta\phi = 0$. This shift, which has been overlooked in previous studies, has a geometric origin arising from the circular symmetry of the $\Delta_{\bf q}$ shell in the FF state. Specifically, in the conventional FF state, the gap landscape $\Delta_{\bf q}$ [e.g., the region with a nonzero $\Delta_{\bf q}$ in the inset of Fig.~\ref{figyc2}(a) for $\gamma_s=0.45\gamma_0$] is nearly degenerate along a circular shell in ${\bf q}$-space, with the optimal pairing momentum ${\bf q}_0$ located near its outer edge. As a result, a supercurrent applied parallel to ${\bf q}_0$ quickly drives the total momentum to the edge of this shell and triggers the gap to collapse, yielding the minimum $J_c$. A current slightly off the antiparallel direction probes more stable segments of the shell,  thereby resulting in the maximum $J_c$. 

For an altermagnetic exchange field, the evolution of the critical current exhibits qualitatively distinct behavior. As the splitting increases from zero, the BCS state [Fig.~\ref{figyc3}(c)] gradually loses its isotropy, and the angular profile of $J_c(\delta\phi)$ starts to develop the  pronounced oscillations. With increasing $\gamma_d$, the critical current along the spin-splitting axes (i.e., the $x$ and $y$ directions: $\delta\phi = \pm\pi/4$ and $\pm3\pi/4$) is progressively suppressed, while the critical current along the nodal directions of the splitting (i.e., the diagonal directions:  $\delta\phi = 0, \pi$ and $\pm\pi/2$) is enhanced. As a result, $J_c(\delta\phi)$ acquires a pronounced fourfold ($C_4$) anisotropy, directly reflecting the $d$-wave character of the altermagnetic spin texture and corresponding to the $C_4$-symmetric, square-like shape of the $\Delta_{\bf q}$ landscape below the critical value [see the insets of Fig.~\ref{figyc2}(b)]. 
   Thus, the maxima of $J_c(\delta\phi)$ occur along the diagonal directions, while the minima lie along $x$ and $y$ axes.     This nontrivial angular misalignment  between the directions of maximal and minimal critical currents gives rise to a {\sl non-collinear} SC diode effect, where the non-reciprocity of the supercurrent is not aligned with a single transport direction, but instead emerges from distinct crystal axes. This  non-collinear SC diode effect is robust, as it originates from the crystal-symmetry-protected nature of the altermagnetic spin splitting, enabling intrinsically stable and symmetry-guided rectification of supercurrents.

With further increase of $\gamma_d$, following the onset of the FF phase, a finite CM momentum ${\bf q}_0$ develops along one of the nodal directions of the altermagnetic spin splitting, and directly breaks SC transport along the opposite direction. This  asymmetry occurs because ${\bf q}_0$ resides near the inner boundary of the gap landscape $\Delta_{\bf q}$, making the system particularly sensitive to supercurrents applied in the $-{\bf q}_0$ direction, which easily push the total momentum beyond the de-pairing threshold and trigger SC breakdown. As a result, only three residual peaks remain in the critical-current profile, aligned with the other nodal directions ($\delta\phi=0,\pm\pi/2$),  since in this early FF regime, the $\Delta_{\bf q}$ landscape still spans a continuous connected region in ${\bf q}$ space [see the inset of Fig.~\ref{figyc2}(b)], although its global maximum has already shifted away from the origin to ${\bf q}_0$.  Notably, as $\gamma_d$ increases further,  the critical-current peaks perpendicular to ${\bf q}_0$ are gradually suppressed. Once the $\Delta_{\bf q}$ landscape fractures into four discrete maxima at $\gamma_d\ge0.63\gamma_0$ [see the inset of Fig.~\ref{figyc2}(b)], the critical-current contour evolves into a broadly peaked unidirectional ($C_1$) pattern aligned with ${\bf q}_0$, i.e., a single preferred SC transport direction is spontaneously selected and pinned. This continuous symmetry reduction in the critical-current profile  from $C_4$ to $C_1$ symmetry, marked by four, then three, and eventually a single peak, is a direct consequence of the smooth-evolution nature of the BCS-FF transition in the $d$-wave altermagnetic system.  

It should be emphasized that the altermagnetism-induced FF state gives rise to a significant genuine and nearly ideal non-reciprocal SC transport behavior, manifested by a pronounced critical-current peak around the ${\bf q}_0$ direction and an entirely vanishing SC response in the reverse, i.e., $J_c(\delta\phi=0)\ne0$ and $J_c(\delta\phi=\pm\pi)=0$.  This non-reciprocity is far more prominent than that in conventional FFLO states driven by isotropic splitting [Fig.~\ref{figyc3}(b)]. In particular, deep in the FF regime, SC transport is sustained only in the vicinity of the ${\bf q}_0$ direction, while all other directions are suppressed, further amplifying the non-reciprocal characteristics and enhancing their relevance for practical implementation. Notably, the altermagnetism-induced FF state does not exhibit any abrupt or substantial SC gap suppression, ensuring the persistence of a robust condensate even in the highly non-reciprocal regime.

\begin{widetext}
\begin{center}
\begin{figure}[htb]
  {\includegraphics[width=17.7cm]{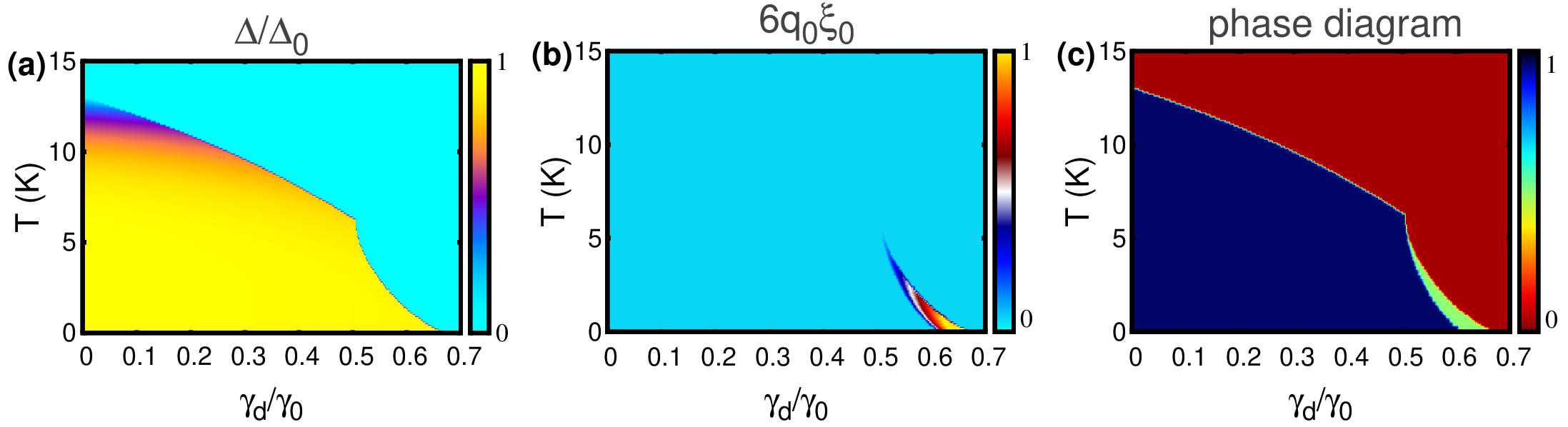}}
  \caption{Temperature–splitting phase diagram of the (a) SC gap and (b) the optimal CM momentum. (c) Overall phase diagram, where the BCS, normal, and FF phases are indicated in blue, red, and green, respectively.} 
\label{Sfigyc2}
\end{figure}
\end{center}
\end{widetext}

{\it Discussion.---}The present study demonstrates that  altermagnetic proximity offers a resilient and field-free route to realizing robust SC diode effects. The combined features revealed, including a non-collinear SC diode effect at weak spin splitting (i.e., in the BCS regime), pronounced SC non-reciprocity at large splitting (i.e., in the FF regime), and unidirectional SC transport deep in the FF state, along with a robust and sizable SC gap throughout the entire SC phase,  make altermagnetic heterostructures promising platforms for symmetry-protected, energy-efficient SC electronics for next-generation quantum and ultra-low-dissipation circuit technologies. {Physically, the gap equation at $q=0$ is an even function of $h_{\bf k}$ and  remains invariant under $C_4$.   One therefore expects that the SC  instability and the resulting critical currents are $C_4$-symmetric. After the system  develops a finite pairing momentum $\mathbf{q}$, the spontaneously selected direction of $\mathbf{q}$  reduces the rotational symmetry from $C_4$ to $C_1$, giving rise to the diode behavior.} 

{While we have focused on low-temperature results to emphasize device performance, the framework can be straightforwardly extended to finite temperatures up to the critical temperature $T_c$. A complete temperature–splitting phase diagram of the order parameters  (the SC gap and the optimal CM momentum) is presented in Fig.~\ref{Sfigyc2}. As shown in the figure, with increasing temperature,  (I) for $\gamma_d \in (0, 0.2\gamma_0)$, the system undergoes a second-order transition from the BCS to the normal state,  characterized by a continuous suppression of the SC gap; (II) for $\gamma_d \in (0.2\gamma_0, 0.5\gamma_0)$, the transition becomes first-order, marked by a discontinuous collapse of the gap, indicative of phase competition or metastability; (III) for $\gamma_d \in (0.5\gamma_0, 0.66\gamma_0)$, the system first undergoes a second-order transition from the BCS to the finite-momentum FF state, followed by a first-order transition from the FF to the normal state, suggesting enhanced sensitivity to thermal fluctuations in the finite-momentum condensate.  A tri-critical Lifshitz point appears at approximately $T \approx 7$~K and $\gamma_d \approx 0.5\gamma_0$, where three distinct phases (BCS, FF, and normal) coexist. This point simultaneously marks (i) the onset of a finite pairing momentum, (ii) the change in the nature of the SC  transition, and (iii) a topology change in the phase diagram. Such a Lifshitz point reflects a qualitative restructuring of the SC ground state, and may manifest experimentally as a nonanalyticity in thermodynamic quantities or anomalous transport behavior near the transition\cite{yang2023optical,yang2024optical}. Moreover, in Supplement Materials~\cite{supple} (see also references~\cite{pekker2015amplitude,PhysRevB.111.174507} therein), we have also derived the emergence of spin-triplet pairing correlations~\cite{PhysRevB.95.075304,PhysRevB.96.134516,yang2018fulde} which may lead to novel phenomena such as an anomalous paramagnetic  Meissner effect, as introduced in Refs.~\cite{RevModPhys.77.1321,RevModPhys.77.935} and experimentally observed in Ref.~\cite{PhysRevX.5.041021}, and cause the magnetoelectric effect~\cite{yang2018fulde,PhysRevLett.118.016802}.}

{\it Acknowledgments.---}This work was supported by the U.S. Department of Energy, Office of Science, Basic Energy Sciences, under Award Number DE-SC0020145, as part of the Computational Materials Sciences Program. F.Y. and L.Q.C. also acknowledge the generous support of Donald W. Hamer Foundation through a Hamer Professorship at Penn State.


\begin{thebibliography}{95}%
\makeatletter
\providecommand \@ifxundefined [1]{%
 \@ifx{#1\undefined}
}%
\providecommand \@ifnum [1]{%
 \ifnum #1\expandafter \@firstoftwo
 \else \expandafter \@secondoftwo
 \fi
}%
\providecommand \@ifx [1]{%
 \ifx #1\expandafter \@firstoftwo
 \else \expandafter \@secondoftwo
 \fi
}%
\providecommand \natexlab [1]{#1}%
\providecommand \enquote  [1]{``#1''}%
\providecommand \bibnamefont  [1]{#1}%
\providecommand \bibfnamefont [1]{#1}%
\providecommand \citenamefont [1]{#1}%
\providecommand \href@noop [0]{\@secondoftwo}%
\providecommand \href [0]{\begingroup \@sanitize@url \@href}%
\providecommand \@href[1]{\@@startlink{#1}\@@href}%
\providecommand \@@href[1]{\endgroup#1\@@endlink}%
\providecommand \@sanitize@url [0]{\catcode `\\12\catcode `\$12\catcode
  `\&12\catcode `\#12\catcode `\^12\catcode `\_12\catcode `\%12\relax}%
\providecommand \@@startlink[1]{}%
\providecommand \@@endlink[0]{}%
\providecommand \url  [0]{\begingroup\@sanitize@url \@url }%
\providecommand \@url [1]{\endgroup\@href {#1}{\urlprefix }}%
\providecommand \urlprefix  [0]{URL }%
\providecommand \Eprint [0]{\href }%
\providecommand \doibase [0]{https://doi.org/}%
\providecommand \selectlanguage [0]{\@gobble}%
\providecommand \bibinfo  [0]{\@secondoftwo}%
\providecommand \bibfield  [0]{\@secondoftwo}%
\providecommand \translation [1]{[#1]}%
\providecommand \BibitemOpen [0]{}%
\providecommand \bibitemStop [0]{}%
\providecommand \bibitemNoStop [0]{.\EOS\space}%
\providecommand \EOS [0]{\spacefactor3000\relax}%
\providecommand \BibitemShut  [1]{\csname bibitem#1\endcsname}%
\let\auto@bib@innerbib\@empty
\bibitem [{\citenamefont {Zhang}\ \emph {et~al.}(2022)\citenamefont {Zhang},
  \citenamefont {Gu}, \citenamefont {Li}, \citenamefont {Hu},\ and\
  \citenamefont {Jiang}}]{PhysRevX.12.041013}%
  \BibitemOpen
  \bibfield  {author} {\bibinfo {author} {\bibfnamefont {Y.}~\bibnamefont
  {Zhang}}, \bibinfo {author} {\bibfnamefont {Y.}~\bibnamefont {Gu}}, \bibinfo
  {author} {\bibfnamefont {P.}~\bibnamefont {Li}}, \bibinfo {author}
  {\bibfnamefont {J.}~\bibnamefont {Hu}},\ and\ \bibinfo {author}
  {\bibfnamefont {K.}~\bibnamefont {Jiang}},\ }\bibfield  {title} {\bibinfo
  {title} {General theory of josephson diodes},\ }\href
  {https://doi.org/10.1103/PhysRevX.12.041013} {\bibfield  {journal} {\bibinfo
  {journal} {Phys. Rev. X}\ }\textbf {\bibinfo {volume} {12}},\ \bibinfo
  {pages} {041013} (\bibinfo {year} {2022})}\BibitemShut {NoStop}%
\bibitem [{\citenamefont {Daido}\ \emph {et~al.}(2022)\citenamefont {Daido},
  \citenamefont {Ikeda},\ and\ \citenamefont
  {Yanase}}]{PhysRevLett.128.037001}%
  \BibitemOpen
  \bibfield  {author} {\bibinfo {author} {\bibfnamefont {A.}~\bibnamefont
  {Daido}}, \bibinfo {author} {\bibfnamefont {Y.}~\bibnamefont {Ikeda}},\ and\
  \bibinfo {author} {\bibfnamefont {Y.}~\bibnamefont {Yanase}},\ }\bibfield
  {title} {\bibinfo {title} {Intrinsic superconducting diode effect},\ }\href
  {https://doi.org/10.1103/PhysRevLett.128.037001} {\bibfield  {journal}
  {\bibinfo  {journal} {Phys. Rev. Lett.}\ }\textbf {\bibinfo {volume} {128}},\
  \bibinfo {pages} {037001} (\bibinfo {year} {2022})}\BibitemShut {NoStop}%
\bibitem [{\citenamefont {Ili\ifmmode~\acute{c}\else \'{c}\fi{}}\ and\
  \citenamefont {Bergeret}(2022)}]{PhysRevLett.128.177001}%
  \BibitemOpen
  \bibfield  {author} {\bibinfo {author} {\bibfnamefont {S.}~\bibnamefont
  {Ili\ifmmode~\acute{c}\else \'{c}\fi{}}}\ and\ \bibinfo {author}
  {\bibfnamefont {F.~S.}\ \bibnamefont {Bergeret}},\ }\bibfield  {title}
  {\bibinfo {title} {Theory of the supercurrent diode effect in rashba
  superconductors with arbitrary disorder},\ }\href
  {https://doi.org/10.1103/PhysRevLett.128.177001} {\bibfield  {journal}
  {\bibinfo  {journal} {Phys. Rev. Lett.}\ }\textbf {\bibinfo {volume} {128}},\
  \bibinfo {pages} {177001} (\bibinfo {year} {2022})}\BibitemShut {NoStop}%
\bibitem [{\citenamefont {Yuan}\ and\ \citenamefont
  {Fu}(2022)}]{doi:10.1073/pnas.2119548119}%
  \BibitemOpen
  \bibfield  {author} {\bibinfo {author} {\bibfnamefont {N.~F.~Q.}\
  \bibnamefont {Yuan}}\ and\ \bibinfo {author} {\bibfnamefont {L.}~\bibnamefont
  {Fu}},\ }\bibfield  {title} {\bibinfo {title} {Supercurrent diode effect and
  finite-momentum superconductors},\ }\href
  {https://doi.org/10.1073/pnas.2119548119} {\bibfield  {journal} {\bibinfo
  {journal} {Proc. Natl. Acad. Sci.}\ }\textbf {\bibinfo {volume} {119}},\
  \bibinfo {pages} {e2119548119} (\bibinfo {year} {2022})}\BibitemShut
  {NoStop}%
\bibitem [{\citenamefont {He}\ \emph {et~al.}(2022)\citenamefont {He},
  \citenamefont {Tanaka},\ and\ \citenamefont {Nagaosa}}]{He_2022}%
  \BibitemOpen
  \bibfield  {author} {\bibinfo {author} {\bibfnamefont {J.~J.}\ \bibnamefont
  {He}}, \bibinfo {author} {\bibfnamefont {Y.}~\bibnamefont {Tanaka}},\ and\
  \bibinfo {author} {\bibfnamefont {N.}~\bibnamefont {Nagaosa}},\ }\bibfield
  {title} {\bibinfo {title} {A phenomenological theory of superconductor
  diodes},\ }\href {https://doi.org/10.1088/1367-2630/ac6766} {\bibfield
  {journal} {\bibinfo  {journal} {New J. Phys.}\ }\textbf {\bibinfo {volume}
  {24}},\ \bibinfo {pages} {053014} (\bibinfo {year} {2022})}\BibitemShut
  {NoStop}%
\bibitem [{\citenamefont {Scammell}\ \emph {et~al.}(2022)\citenamefont
  {Scammell}, \citenamefont {Li},\ and\ \citenamefont
  {Scheurer}}]{Scammell_2022}%
  \BibitemOpen
  \bibfield  {author} {\bibinfo {author} {\bibfnamefont {H.~D.}\ \bibnamefont
  {Scammell}}, \bibinfo {author} {\bibfnamefont {J.~I.~A.}\ \bibnamefont
  {Li}},\ and\ \bibinfo {author} {\bibfnamefont {M.~S.}\ \bibnamefont
  {Scheurer}},\ }\bibfield  {title} {\bibinfo {title} {Theory of zero-field
  superconducting diode effect in twisted trilayer graphene},\ }\href
  {https://doi.org/10.1088/2053-1583/ac5b16} {\bibfield  {journal} {\bibinfo
  {journal} {2D Mater.}\ }\textbf {\bibinfo {volume} {9}},\ \bibinfo {pages}
  {025027} (\bibinfo {year} {2022})}\BibitemShut {NoStop}%
\bibitem [{\citenamefont {Davydova}\ \emph {et~al.}(2022)\citenamefont
  {Davydova}, \citenamefont {Prembabu},\ and\ \citenamefont
  {Fu}}]{doi:10.1126/sciadv.abo0309}%
  \BibitemOpen
  \bibfield  {author} {\bibinfo {author} {\bibfnamefont {M.}~\bibnamefont
  {Davydova}}, \bibinfo {author} {\bibfnamefont {S.}~\bibnamefont {Prembabu}},\
  and\ \bibinfo {author} {\bibfnamefont {L.}~\bibnamefont {Fu}},\ }\bibfield
  {title} {\bibinfo {title} {Universal josephson diode effect},\ }\href
  {https://doi.org/10.1126/sciadv.abo0309} {\bibfield  {journal} {\bibinfo
  {journal} {Sci. Adv.}\ }\textbf {\bibinfo {volume} {8}},\ \bibinfo {pages}
  {eabo0309} (\bibinfo {year} {2022})}\BibitemShut {NoStop}%
\bibitem [{\citenamefont {Tanaka}\ \emph {et~al.}(2022)\citenamefont {Tanaka},
  \citenamefont {Lu},\ and\ \citenamefont {Nagaosa}}]{PhysRevB.106.214524}%
  \BibitemOpen
  \bibfield  {author} {\bibinfo {author} {\bibfnamefont {Y.}~\bibnamefont
  {Tanaka}}, \bibinfo {author} {\bibfnamefont {B.}~\bibnamefont {Lu}},\ and\
  \bibinfo {author} {\bibfnamefont {N.}~\bibnamefont {Nagaosa}},\ }\bibfield
  {title} {\bibinfo {title} {Theory of giant diode effect in $d$-wave
  superconductor junctions on the surface of a topological insulator},\ }\href
  {https://doi.org/10.1103/PhysRevB.106.214524} {\bibfield  {journal} {\bibinfo
   {journal} {Phys. Rev. B}\ }\textbf {\bibinfo {volume} {106}},\ \bibinfo
  {pages} {214524} (\bibinfo {year} {2022})}\BibitemShut {NoStop}%
\bibitem [{\citenamefont {Bauriedl}\ \emph {et~al.}(2022)\citenamefont
  {Bauriedl}, \citenamefont {B{\"a}uml}, \citenamefont {Fuchs}, \citenamefont
  {Baumgartner}, \citenamefont {Paulik}, \citenamefont {Bauer}, \citenamefont
  {Lin}, \citenamefont {Lupton}, \citenamefont {Taniguchi}, \citenamefont
  {Watanabe} \emph {et~al.}}]{bauriedl2022supercurrent}%
  \BibitemOpen
  \bibfield  {author} {\bibinfo {author} {\bibfnamefont {L.}~\bibnamefont
  {Bauriedl}}, \bibinfo {author} {\bibfnamefont {C.}~\bibnamefont {B{\"a}uml}},
  \bibinfo {author} {\bibfnamefont {L.}~\bibnamefont {Fuchs}}, \bibinfo
  {author} {\bibfnamefont {C.}~\bibnamefont {Baumgartner}}, \bibinfo {author}
  {\bibfnamefont {N.}~\bibnamefont {Paulik}}, \bibinfo {author} {\bibfnamefont
  {J.~M.}\ \bibnamefont {Bauer}}, \bibinfo {author} {\bibfnamefont {K.-Q.}\
  \bibnamefont {Lin}}, \bibinfo {author} {\bibfnamefont {J.~M.}\ \bibnamefont
  {Lupton}}, \bibinfo {author} {\bibfnamefont {T.}~\bibnamefont {Taniguchi}},
  \bibinfo {author} {\bibfnamefont {K.}~\bibnamefont {Watanabe}}, \emph
  {et~al.},\ }\bibfield  {title} {\bibinfo {title} {Supercurrent diode effect
  and magnetochiral anisotropy in few-layer nbse2},\ }\href@noop {} {\bibfield
  {journal} {\bibinfo  {journal} {Nat. Commun.}\ }\textbf {\bibinfo {volume}
  {13}},\ \bibinfo {pages} {4266} (\bibinfo {year} {2022})}\BibitemShut
  {NoStop}%
\bibitem [{\citenamefont {Ando}\ \emph {et~al.}(2020)\citenamefont {Ando},
  \citenamefont {Miyasaka}, \citenamefont {Li}, \citenamefont {Ishizuka},
  \citenamefont {Arakawa}, \citenamefont {Shiota}, \citenamefont {Moriyama},
  \citenamefont {Yanase},\ and\ \citenamefont {Ono}}]{ando2020observation}%
  \BibitemOpen
  \bibfield  {author} {\bibinfo {author} {\bibfnamefont {F.}~\bibnamefont
  {Ando}}, \bibinfo {author} {\bibfnamefont {Y.}~\bibnamefont {Miyasaka}},
  \bibinfo {author} {\bibfnamefont {T.}~\bibnamefont {Li}}, \bibinfo {author}
  {\bibfnamefont {J.}~\bibnamefont {Ishizuka}}, \bibinfo {author}
  {\bibfnamefont {T.}~\bibnamefont {Arakawa}}, \bibinfo {author} {\bibfnamefont
  {Y.}~\bibnamefont {Shiota}}, \bibinfo {author} {\bibfnamefont
  {T.}~\bibnamefont {Moriyama}}, \bibinfo {author} {\bibfnamefont
  {Y.}~\bibnamefont {Yanase}},\ and\ \bibinfo {author} {\bibfnamefont
  {T.}~\bibnamefont {Ono}},\ }\bibfield  {title} {\bibinfo {title} {Observation
  of superconducting diode effect},\ }\href@noop {} {\bibfield  {journal}
  {\bibinfo  {journal} {Nature}\ }\textbf {\bibinfo {volume} {584}},\ \bibinfo
  {pages} {373} (\bibinfo {year} {2020})}\BibitemShut {NoStop}%
\bibitem [{\citenamefont {Wu}\ \emph {et~al.}(2022)\citenamefont {Wu},
  \citenamefont {Wang}, \citenamefont {Xu}, \citenamefont {Sivakumar},
  \citenamefont {Pasco}, \citenamefont {Filippozzi}, \citenamefont {Parkin},
  \citenamefont {Zeng}, \citenamefont {McQueen},\ and\ \citenamefont
  {Ali}}]{wu2022field}%
  \BibitemOpen
  \bibfield  {author} {\bibinfo {author} {\bibfnamefont {H.}~\bibnamefont
  {Wu}}, \bibinfo {author} {\bibfnamefont {Y.}~\bibnamefont {Wang}}, \bibinfo
  {author} {\bibfnamefont {Y.}~\bibnamefont {Xu}}, \bibinfo {author}
  {\bibfnamefont {P.~K.}\ \bibnamefont {Sivakumar}}, \bibinfo {author}
  {\bibfnamefont {C.}~\bibnamefont {Pasco}}, \bibinfo {author} {\bibfnamefont
  {U.}~\bibnamefont {Filippozzi}}, \bibinfo {author} {\bibfnamefont {S.~S.}\
  \bibnamefont {Parkin}}, \bibinfo {author} {\bibfnamefont {Y.-J.}\
  \bibnamefont {Zeng}}, \bibinfo {author} {\bibfnamefont {T.}~\bibnamefont
  {McQueen}},\ and\ \bibinfo {author} {\bibfnamefont {M.~N.}\ \bibnamefont
  {Ali}},\ }\bibfield  {title} {\bibinfo {title} {The field-free josephson
  diode in a van der waals heterostructure},\ }\href@noop {} {\bibfield
  {journal} {\bibinfo  {journal} {Nature}\ }\textbf {\bibinfo {volume} {604}},\
  \bibinfo {pages} {653} (\bibinfo {year} {2022})}\BibitemShut {NoStop}%
\bibitem [{\citenamefont {Wakatsuki}\ and\ \citenamefont
  {Nagaosa}(2018)}]{PhysRevLett.121.026601}%
  \BibitemOpen
  \bibfield  {author} {\bibinfo {author} {\bibfnamefont {R.}~\bibnamefont
  {Wakatsuki}}\ and\ \bibinfo {author} {\bibfnamefont {N.}~\bibnamefont
  {Nagaosa}},\ }\bibfield  {title} {\bibinfo {title} {Nonreciprocal current in
  noncentrosymmetric rashba superconductors},\ }\href
  {https://doi.org/10.1103/PhysRevLett.121.026601} {\bibfield  {journal}
  {\bibinfo  {journal} {Phys. Rev. Lett.}\ }\textbf {\bibinfo {volume} {121}},\
  \bibinfo {pages} {026601} (\bibinfo {year} {2018})}\BibitemShut {NoStop}%
\bibitem [{\citenamefont {Daido}\ and\ \citenamefont
  {Yanase}(2022)}]{PhysRevB.106.205206}%
  \BibitemOpen
  \bibfield  {author} {\bibinfo {author} {\bibfnamefont {A.}~\bibnamefont
  {Daido}}\ and\ \bibinfo {author} {\bibfnamefont {Y.}~\bibnamefont {Yanase}},\
  }\bibfield  {title} {\bibinfo {title} {Superconducting diode effect and
  nonreciprocal transition lines},\ }\href
  {https://doi.org/10.1103/PhysRevB.106.205206} {\bibfield  {journal} {\bibinfo
   {journal} {Phys. Rev. B}\ }\textbf {\bibinfo {volume} {106}},\ \bibinfo
  {pages} {205206} (\bibinfo {year} {2022})}\BibitemShut {NoStop}%
\bibitem [{\citenamefont {Nagaosa}\ and\ \citenamefont
  {Yanase}(2024)}]{nagaosa2024nonreciprocal}%
  \BibitemOpen
  \bibfield  {author} {\bibinfo {author} {\bibfnamefont {N.}~\bibnamefont
  {Nagaosa}}\ and\ \bibinfo {author} {\bibfnamefont {Y.}~\bibnamefont
  {Yanase}},\ }\bibfield  {title} {\bibinfo {title} {Nonreciprocal transport
  and optical phenomena in quantum materials},\ }\href@noop {} {\bibfield
  {journal} {\bibinfo  {journal} {Annu. Rev. Condens. Matter Phys.}\ }\textbf
  {\bibinfo {volume} {15}},\ \bibinfo {pages} {63} (\bibinfo {year}
  {2024})}\BibitemShut {NoStop}%
\bibitem [{\citenamefont {Larkin}\ and\ \citenamefont
  {Ovchinnikov}(1965)}]{larkin1965zh}%
  \BibitemOpen
  \bibfield  {author} {\bibinfo {author} {\bibfnamefont {A.}~\bibnamefont
  {Larkin}}\ and\ \bibinfo {author} {\bibfnamefont {Y.~N.}\ \bibnamefont
  {Ovchinnikov}},\ }\bibfield  {title} {\bibinfo {title} {Nonuniform state of
  superconductors},\ }\href@noop {} {\bibfield  {journal} {\bibinfo  {journal}
  {JETP}\ }\textbf {\bibinfo {volume} {20}},\ \bibinfo {pages} {762} (\bibinfo
  {year} {1965})}\BibitemShut {NoStop}%
\bibitem [{\citenamefont {Fulde}\ and\ \citenamefont
  {Ferrell}(1964)}]{fulde1964superconductivity}%
  \BibitemOpen
  \bibfield  {author} {\bibinfo {author} {\bibfnamefont {P.}~\bibnamefont
  {Fulde}}\ and\ \bibinfo {author} {\bibfnamefont {R.~A.}\ \bibnamefont
  {Ferrell}},\ }\bibfield  {title} {\bibinfo {title} {Superconductivity in a
  strong spin-exchange field},\ }\href@noop {} {\bibfield  {journal} {\bibinfo
  {journal} {Phys. Rev.}\ }\textbf {\bibinfo {volume} {135}},\ \bibinfo {pages}
  {A550} (\bibinfo {year} {1964})}\BibitemShut {NoStop}%
\bibitem [{\citenamefont {Abrikosov}\ \emph {et~al.}(2012)\citenamefont
  {Abrikosov}, \citenamefont {Gorkov},\ and\ \citenamefont
  {Dzyaloshinski}}]{abrikosov2012methods}%
  \BibitemOpen
  \bibfield  {author} {\bibinfo {author} {\bibfnamefont {A.~A.}\ \bibnamefont
  {Abrikosov}}, \bibinfo {author} {\bibfnamefont {L.~P.}\ \bibnamefont
  {Gorkov}},\ and\ \bibinfo {author} {\bibfnamefont {I.~E.}\ \bibnamefont
  {Dzyaloshinski}},\ }\href@noop {} {\emph {\bibinfo {title} {Methods of
  quantum field theory in statistical physics}}}\ (\bibinfo  {publisher}
  {Courier Corporation},\ \bibinfo {year} {2012})\BibitemShut {NoStop}%
\bibitem [{\citenamefont {Goldstone}(1961)}]{goldstone1961field}%
  \BibitemOpen
  \bibfield  {author} {\bibinfo {author} {\bibfnamefont {J.}~\bibnamefont
  {Goldstone}},\ }\bibfield  {title} {\bibinfo {title} {Field theories with
  superconductor solutions},\ }\href@noop {} {\bibfield  {journal} {\bibinfo
  {journal} {Il Nuovo Cimento}\ }\textbf {\bibinfo {volume} {19}},\ \bibinfo
  {pages} {154} (\bibinfo {year} {1961})}\BibitemShut {NoStop}%
\bibitem [{\citenamefont {Goldstone}\ \emph {et~al.}(1962)\citenamefont
  {Goldstone}, \citenamefont {Salam},\ and\ \citenamefont
  {Weinberg}}]{goldstone1962broken}%
  \BibitemOpen
  \bibfield  {author} {\bibinfo {author} {\bibfnamefont {J.}~\bibnamefont
  {Goldstone}}, \bibinfo {author} {\bibfnamefont {A.}~\bibnamefont {Salam}},\
  and\ \bibinfo {author} {\bibfnamefont {S.}~\bibnamefont {Weinberg}},\
  }\bibfield  {title} {\bibinfo {title} {Broken symmetries},\ }\href@noop {}
  {\bibfield  {journal} {\bibinfo  {journal} {Phys. Rev.}\ }\textbf {\bibinfo
  {volume} {127}},\ \bibinfo {pages} {965} (\bibinfo {year}
  {1962})}\BibitemShut {NoStop}%
\bibitem [{\citenamefont {Yang}\ and\ \citenamefont
  {Wu}(2018{\natexlab{a}})}]{yang2018fulde}%
  \BibitemOpen
  \bibfield  {author} {\bibinfo {author} {\bibfnamefont {F.}~\bibnamefont
  {Yang}}\ and\ \bibinfo {author} {\bibfnamefont {M.~W.}\ \bibnamefont {Wu}},\
  }\bibfield  {title} {\bibinfo {title} {{Fulde--Ferrell} state in
  spin--orbit-coupled superconductor: Application to {Dresselhaus SOC}},\
  }\href@noop {} {\bibfield  {journal} {\bibinfo  {journal} {J. Low Temp.
  Phys.}\ }\textbf {\bibinfo {volume} {192}},\ \bibinfo {pages} {241} (\bibinfo
  {year} {2018}{\natexlab{a}})}\BibitemShut {NoStop}%
\bibitem [{\citenamefont {Dong}\ \emph {et~al.}(2013)\citenamefont {Dong},
  \citenamefont {Jiang},\ and\ \citenamefont {Pu}}]{Dong_2013}%
  \BibitemOpen
  \bibfield  {author} {\bibinfo {author} {\bibfnamefont {L.}~\bibnamefont
  {Dong}}, \bibinfo {author} {\bibfnamefont {L.}~\bibnamefont {Jiang}},\ and\
  \bibinfo {author} {\bibfnamefont {H.}~\bibnamefont {Pu}},\ }\bibfield
  {title} {\bibinfo {title} {Fulde–ferrell pairing instability in
  spin–orbit coupled fermi gas},\ }\href
  {https://doi.org/10.1088/1367-2630/15/7/075014} {\bibfield  {journal}
  {\bibinfo  {journal} {New J. Phys.}\ }\textbf {\bibinfo {volume} {15}},\
  \bibinfo {pages} {075014} (\bibinfo {year} {2013})}\BibitemShut {NoStop}%
\bibitem [{\citenamefont {Xu}\ \emph {et~al.}(2014)\citenamefont {Xu},
  \citenamefont {Qu}, \citenamefont {Gong},\ and\ \citenamefont
  {Zhang}}]{PhysRevA.89.013607}%
  \BibitemOpen
  \bibfield  {author} {\bibinfo {author} {\bibfnamefont {Y.}~\bibnamefont
  {Xu}}, \bibinfo {author} {\bibfnamefont {C.}~\bibnamefont {Qu}}, \bibinfo
  {author} {\bibfnamefont {M.}~\bibnamefont {Gong}},\ and\ \bibinfo {author}
  {\bibfnamefont {C.}~\bibnamefont {Zhang}},\ }\bibfield  {title} {\bibinfo
  {title} {Competing superfluid orders in spin-orbit-coupled fermionic
  cold-atom optical lattices},\ }\href
  {https://doi.org/10.1103/PhysRevA.89.013607} {\bibfield  {journal} {\bibinfo
  {journal} {Phys. Rev. A}\ }\textbf {\bibinfo {volume} {89}},\ \bibinfo
  {pages} {013607} (\bibinfo {year} {2014})}\BibitemShut {NoStop}%
\bibitem [{\citenamefont {Barzykin}\ and\ \citenamefont
  {Gor'kov}(2002)}]{PhysRevLett.89.227002}%
  \BibitemOpen
  \bibfield  {author} {\bibinfo {author} {\bibfnamefont {V.}~\bibnamefont
  {Barzykin}}\ and\ \bibinfo {author} {\bibfnamefont {L.~P.}\ \bibnamefont
  {Gor'kov}},\ }\bibfield  {title} {\bibinfo {title} {Inhomogeneous stripe
  phase revisited for surface superconductivity},\ }\href
  {https://doi.org/10.1103/PhysRevLett.89.227002} {\bibfield  {journal}
  {\bibinfo  {journal} {Phys. Rev. Lett.}\ }\textbf {\bibinfo {volume} {89}},\
  \bibinfo {pages} {227002} (\bibinfo {year} {2002})}\BibitemShut {NoStop}%
\bibitem [{\citenamefont {Dimitrova}\ and\ \citenamefont
  {Feigel'man}(2007)}]{PhysRevB.76.014522}%
  \BibitemOpen
  \bibfield  {author} {\bibinfo {author} {\bibfnamefont {O.}~\bibnamefont
  {Dimitrova}}\ and\ \bibinfo {author} {\bibfnamefont {M.~V.}\ \bibnamefont
  {Feigel'man}},\ }\bibfield  {title} {\bibinfo {title} {Theory of a
  two-dimensional superconductor with broken inversion symmetry},\ }\href
  {https://doi.org/10.1103/PhysRevB.76.014522} {\bibfield  {journal} {\bibinfo
  {journal} {Phys. Rev. B}\ }\textbf {\bibinfo {volume} {76}},\ \bibinfo
  {pages} {014522} (\bibinfo {year} {2007})}\BibitemShut {NoStop}%
\bibitem [{\citenamefont {Agterberg}\ and\ \citenamefont
  {Kaur}(2007)}]{PhysRevB.75.064511}%
  \BibitemOpen
  \bibfield  {author} {\bibinfo {author} {\bibfnamefont {D.~F.}\ \bibnamefont
  {Agterberg}}\ and\ \bibinfo {author} {\bibfnamefont {R.~P.}\ \bibnamefont
  {Kaur}},\ }\bibfield  {title} {\bibinfo {title} {Magnetic-field-induced
  helical and stripe phases in rashba superconductors},\ }\href
  {https://doi.org/10.1103/PhysRevB.75.064511} {\bibfield  {journal} {\bibinfo
  {journal} {Phys. Rev. B}\ }\textbf {\bibinfo {volume} {75}},\ \bibinfo
  {pages} {064511} (\bibinfo {year} {2007})}\BibitemShut {NoStop}%
\bibitem [{\citenamefont {Xu}\ and\ \citenamefont
  {Zhang}(2015)}]{PhysRevLett.114.110401}%
  \BibitemOpen
  \bibfield  {author} {\bibinfo {author} {\bibfnamefont {Y.}~\bibnamefont
  {Xu}}\ and\ \bibinfo {author} {\bibfnamefont {C.}~\bibnamefont {Zhang}},\
  }\bibfield  {title} {\bibinfo {title} {Berezinskii-kosterlitz-thouless phase
  transition in 2d spin-orbit-coupled fulde-ferrell superfluids},\ }\href
  {https://doi.org/10.1103/PhysRevLett.114.110401} {\bibfield  {journal}
  {\bibinfo  {journal} {Phys. Rev. Lett.}\ }\textbf {\bibinfo {volume} {114}},\
  \bibinfo {pages} {110401} (\bibinfo {year} {2015})}\BibitemShut {NoStop}%
\bibitem [{\citenamefont {Hong}\ \emph {et~al.}(2025)\citenamefont {Hong},
  \citenamefont {Park},\ and\ \citenamefont {Kim}}]{PhysRevB.111.054501}%
  \BibitemOpen
  \bibfield  {author} {\bibinfo {author} {\bibfnamefont {S.}~\bibnamefont
  {Hong}}, \bibinfo {author} {\bibfnamefont {M.~J.}\ \bibnamefont {Park}},\
  and\ \bibinfo {author} {\bibfnamefont {K.-M.}\ \bibnamefont {Kim}},\
  }\bibfield  {title} {\bibinfo {title} {Unconventional $p$-wave and
  finite-momentum superconductivity induced by altermagnetism through the
  formation of bogoliubov fermi surface},\ }\href
  {https://doi.org/10.1103/PhysRevB.111.054501} {\bibfield  {journal} {\bibinfo
   {journal} {Phys. Rev. B}\ }\textbf {\bibinfo {volume} {111}},\ \bibinfo
  {pages} {054501} (\bibinfo {year} {2025})}\BibitemShut {NoStop}%
\bibitem [{\citenamefont {{\v{S}}mejkal}\ \emph
  {et~al.}(2022{\natexlab{a}})\citenamefont {{\v{S}}mejkal}, \citenamefont
  {Sinova},\ and\ \citenamefont {Jungwirth}}]{vsmejkal2022beyond}%
  \BibitemOpen
  \bibfield  {author} {\bibinfo {author} {\bibfnamefont {L.}~\bibnamefont
  {{\v{S}}mejkal}}, \bibinfo {author} {\bibfnamefont {J.}~\bibnamefont
  {Sinova}},\ and\ \bibinfo {author} {\bibfnamefont {T.}~\bibnamefont
  {Jungwirth}},\ }\bibfield  {title} {\bibinfo {title} {Beyond conventional
  ferromagnetism and antiferromagnetism: A phase with nonrelativistic spin and
  crystal rotation symmetry},\ }\href@noop {} {\bibfield  {journal} {\bibinfo
  {journal} {Phys. Rev. X.}\ }\textbf {\bibinfo {volume} {12}},\ \bibinfo
  {pages} {031042} (\bibinfo {year} {2022}{\natexlab{a}})}\BibitemShut
  {NoStop}%
\bibitem [{\citenamefont {{\v{S}}mejkal}\ \emph
  {et~al.}(2022{\natexlab{b}})\citenamefont {{\v{S}}mejkal}, \citenamefont
  {Sinova},\ and\ \citenamefont {Jungwirth}}]{vsmejkal2022emerging}%
  \BibitemOpen
  \bibfield  {author} {\bibinfo {author} {\bibfnamefont {L.}~\bibnamefont
  {{\v{S}}mejkal}}, \bibinfo {author} {\bibfnamefont {J.}~\bibnamefont
  {Sinova}},\ and\ \bibinfo {author} {\bibfnamefont {T.}~\bibnamefont
  {Jungwirth}},\ }\bibfield  {title} {\bibinfo {title} {Emerging research
  landscape of altermagnetism},\ }\href@noop {} {\bibfield  {journal} {\bibinfo
   {journal} {Phys. Rev. X.}\ }\textbf {\bibinfo {volume} {12}},\ \bibinfo
  {pages} {040501} (\bibinfo {year} {2022}{\natexlab{b}})}\BibitemShut
  {NoStop}%
\bibitem [{\citenamefont {Bhowal}\ and\ \citenamefont
  {Spaldin}(2024)}]{bhowal2024ferroically}%
  \BibitemOpen
  \bibfield  {author} {\bibinfo {author} {\bibfnamefont {S.}~\bibnamefont
  {Bhowal}}\ and\ \bibinfo {author} {\bibfnamefont {N.~A.}\ \bibnamefont
  {Spaldin}},\ }\bibfield  {title} {\bibinfo {title} {Ferroically ordered
  magnetic octupoles in d-wave altermagnets},\ }\href@noop {} {\bibfield
  {journal} {\bibinfo  {journal} {Phys. Rev. X.}\ }\textbf {\bibinfo {volume}
  {14}},\ \bibinfo {pages} {011019} (\bibinfo {year} {2024})}\BibitemShut
  {NoStop}%
\bibitem [{\citenamefont {Reichlova}\ \emph {et~al.}(2024)\citenamefont
  {Reichlova}, \citenamefont {L.~Seeger}, \citenamefont {G.~Hern{\'a}ndez},
  \citenamefont {Kounta}, \citenamefont {Schlitz}, \citenamefont {Kriegner},
  \citenamefont {Ritzinger}, \citenamefont {Lammel}, \citenamefont
  {Leivisk{\"a}}, \citenamefont {B.~Hellenes} \emph
  {et~al.}}]{reichlova2024observation}%
  \BibitemOpen
  \bibfield  {author} {\bibinfo {author} {\bibfnamefont {H.}~\bibnamefont
  {Reichlova}}, \bibinfo {author} {\bibfnamefont {R.}~\bibnamefont
  {L.~Seeger}}, \bibinfo {author} {\bibfnamefont {R.}~\bibnamefont
  {G.~Hern{\'a}ndez}}, \bibinfo {author} {\bibfnamefont {I.}~\bibnamefont
  {Kounta}}, \bibinfo {author} {\bibfnamefont {R.}~\bibnamefont {Schlitz}},
  \bibinfo {author} {\bibfnamefont {D.}~\bibnamefont {Kriegner}}, \bibinfo
  {author} {\bibfnamefont {P.}~\bibnamefont {Ritzinger}}, \bibinfo {author}
  {\bibfnamefont {M.}~\bibnamefont {Lammel}}, \bibinfo {author} {\bibfnamefont
  {M.}~\bibnamefont {Leivisk{\"a}}}, \bibinfo {author} {\bibfnamefont
  {A.}~\bibnamefont {B.~Hellenes}}, \emph {et~al.},\ }\bibfield  {title}
  {\bibinfo {title} {Observation of a spontaneous anomalous hall response in
  the {${\mathrm{Mn}}_{5}$${\mathrm{Si}}_{3}$} d-wave altermagnet candidate},\
  }\href@noop {} {\bibfield  {journal} {\bibinfo  {journal} {Nat. Commun.}\
  }\textbf {\bibinfo {volume} {15}},\ \bibinfo {pages} {4961} (\bibinfo {year}
  {2024})}\BibitemShut {NoStop}%
\bibitem [{\citenamefont {Ding}\ \emph {et~al.}(2024)\citenamefont {Ding},
  \citenamefont {Jiang}, \citenamefont {Chen}, \citenamefont {Tao},
  \citenamefont {Liu}, \citenamefont {Li}, \citenamefont {Liu}, \citenamefont
  {Sun}, \citenamefont {Cheng}, \citenamefont {Liu} \emph
  {et~al.}}]{ding2024large}%
  \BibitemOpen
  \bibfield  {author} {\bibinfo {author} {\bibfnamefont {J.}~\bibnamefont
  {Ding}}, \bibinfo {author} {\bibfnamefont {Z.}~\bibnamefont {Jiang}},
  \bibinfo {author} {\bibfnamefont {X.}~\bibnamefont {Chen}}, \bibinfo {author}
  {\bibfnamefont {Z.}~\bibnamefont {Tao}}, \bibinfo {author} {\bibfnamefont
  {Z.}~\bibnamefont {Liu}}, \bibinfo {author} {\bibfnamefont {T.}~\bibnamefont
  {Li}}, \bibinfo {author} {\bibfnamefont {J.}~\bibnamefont {Liu}}, \bibinfo
  {author} {\bibfnamefont {J.}~\bibnamefont {Sun}}, \bibinfo {author}
  {\bibfnamefont {J.}~\bibnamefont {Cheng}}, \bibinfo {author} {\bibfnamefont
  {J.}~\bibnamefont {Liu}}, \emph {et~al.},\ }\bibfield  {title} {\bibinfo
  {title} {Large band splitting in g-wave altermagnet crsb},\ }\href@noop {}
  {\bibfield  {journal} {\bibinfo  {journal} {Phys. Rev. Lett.}\ }\textbf
  {\bibinfo {volume} {133}},\ \bibinfo {pages} {206401} (\bibinfo {year}
  {2024})}\BibitemShut {NoStop}%
\bibitem [{\citenamefont {Liu}\ \emph {et~al.}(2024)\citenamefont {Liu},
  \citenamefont {Ozeki}, \citenamefont {Asai}, \citenamefont {Itoh},\ and\
  \citenamefont {Masuda}}]{PhysRevLett.133.156702}%
  \BibitemOpen
  \bibfield  {author} {\bibinfo {author} {\bibfnamefont {Z.}~\bibnamefont
  {Liu}}, \bibinfo {author} {\bibfnamefont {M.}~\bibnamefont {Ozeki}}, \bibinfo
  {author} {\bibfnamefont {S.}~\bibnamefont {Asai}}, \bibinfo {author}
  {\bibfnamefont {S.}~\bibnamefont {Itoh}},\ and\ \bibinfo {author}
  {\bibfnamefont {T.}~\bibnamefont {Masuda}},\ }\bibfield  {title} {\bibinfo
  {title} {Chiral split magnon in altermagnetic mnte},\ }\href@noop {}
  {\bibfield  {journal} {\bibinfo  {journal} {Phys. Rev. Lett.}\ }\textbf
  {\bibinfo {volume} {133}},\ \bibinfo {pages} {156702} (\bibinfo {year}
  {2024})}\BibitemShut {NoStop}%
\bibitem [{\citenamefont {Zhang}\ \emph {et~al.}(2025)\citenamefont {Zhang},
  \citenamefont {Cheng}, \citenamefont {Yin}, \citenamefont {Liu},
  \citenamefont {Deng}, \citenamefont {Qiao}, \citenamefont {Shi},
  \citenamefont {Zhang}, \citenamefont {Lin}, \citenamefont {Liu} \emph
  {et~al.}}]{zhang2025crystal}%
  \BibitemOpen
  \bibfield  {author} {\bibinfo {author} {\bibfnamefont {F.}~\bibnamefont
  {Zhang}}, \bibinfo {author} {\bibfnamefont {X.}~\bibnamefont {Cheng}},
  \bibinfo {author} {\bibfnamefont {Z.}~\bibnamefont {Yin}}, \bibinfo {author}
  {\bibfnamefont {C.}~\bibnamefont {Liu}}, \bibinfo {author} {\bibfnamefont
  {L.}~\bibnamefont {Deng}}, \bibinfo {author} {\bibfnamefont {Y.}~\bibnamefont
  {Qiao}}, \bibinfo {author} {\bibfnamefont {Z.}~\bibnamefont {Shi}}, \bibinfo
  {author} {\bibfnamefont {S.}~\bibnamefont {Zhang}}, \bibinfo {author}
  {\bibfnamefont {J.}~\bibnamefont {Lin}}, \bibinfo {author} {\bibfnamefont
  {Z.}~\bibnamefont {Liu}}, \emph {et~al.},\ }\bibfield  {title} {\bibinfo
  {title} {Crystal-symmetry-paired spin--valley locking in a layered
  room-temperature metallic altermagnet candidate},\ }\href@noop {} {\bibfield
  {journal} {\bibinfo  {journal} {Nat. Phys.}\ }\textbf {\bibinfo {volume}
  {21}},\ \bibinfo {pages} {760} (\bibinfo {year} {2025})}\BibitemShut
  {NoStop}%
\bibitem [{\citenamefont {Lin}\ \emph {et~al.}(2024)\citenamefont {Lin},
  \citenamefont {Chen}, \citenamefont {Lu}, \citenamefont {Liang},
  \citenamefont {Feng}, \citenamefont {Yamagami}, \citenamefont {Osiecki},
  \citenamefont {Leandersson}, \citenamefont {Thiagarajan}, \citenamefont {Liu}
  \emph {et~al.}}]{lin2024observation}%
  \BibitemOpen
  \bibfield  {author} {\bibinfo {author} {\bibfnamefont {Z.}~\bibnamefont
  {Lin}}, \bibinfo {author} {\bibfnamefont {D.}~\bibnamefont {Chen}}, \bibinfo
  {author} {\bibfnamefont {W.}~\bibnamefont {Lu}}, \bibinfo {author}
  {\bibfnamefont {X.}~\bibnamefont {Liang}}, \bibinfo {author} {\bibfnamefont
  {S.}~\bibnamefont {Feng}}, \bibinfo {author} {\bibfnamefont {K.}~\bibnamefont
  {Yamagami}}, \bibinfo {author} {\bibfnamefont {J.}~\bibnamefont {Osiecki}},
  \bibinfo {author} {\bibfnamefont {M.}~\bibnamefont {Leandersson}}, \bibinfo
  {author} {\bibfnamefont {B.}~\bibnamefont {Thiagarajan}}, \bibinfo {author}
  {\bibfnamefont {J.}~\bibnamefont {Liu}}, \emph {et~al.},\ }\bibfield  {title}
  {\bibinfo {title} {Observation of giant spin splitting and d-wave spin
  texture in room temperature altermagnet {${\mathrm{RuO}}_{2}$}},\ }\href@noop
  {} {\bibfield  {journal} {\bibinfo  {journal} {arXiv:2402.04995}\ } (\bibinfo
  {year} {2024})}\BibitemShut {NoStop}%
\bibitem [{\citenamefont {Bai}\ \emph {et~al.}(2024)\citenamefont {Bai},
  \citenamefont {Feng}, \citenamefont {Liu}, \citenamefont {{\v{S}}mejkal},
  \citenamefont {Mokrousov},\ and\ \citenamefont
  {Yao}}]{bai2024altermagnetism}%
  \BibitemOpen
  \bibfield  {author} {\bibinfo {author} {\bibfnamefont {L.}~\bibnamefont
  {Bai}}, \bibinfo {author} {\bibfnamefont {W.}~\bibnamefont {Feng}}, \bibinfo
  {author} {\bibfnamefont {S.}~\bibnamefont {Liu}}, \bibinfo {author}
  {\bibfnamefont {L.}~\bibnamefont {{\v{S}}mejkal}}, \bibinfo {author}
  {\bibfnamefont {Y.}~\bibnamefont {Mokrousov}},\ and\ \bibinfo {author}
  {\bibfnamefont {Y.}~\bibnamefont {Yao}},\ }\bibfield  {title} {\bibinfo
  {title} {Altermagnetism: Exploring new frontiers in magnetism and
  spintronics},\ }\href@noop {} {\bibfield  {journal} {\bibinfo  {journal}
  {Adv. Funct. Mater.}\ }\textbf {\bibinfo {volume} {34}},\ \bibinfo {pages}
  {2409327} (\bibinfo {year} {2024})}\BibitemShut {NoStop}%
\bibitem [{\citenamefont {Wei{\ss}enhofer}\ and\ \citenamefont
  {Marmodoro}(2024)}]{weissenhofer2024atomistic}%
  \BibitemOpen
  \bibfield  {author} {\bibinfo {author} {\bibfnamefont {M.}~\bibnamefont
  {Wei{\ss}enhofer}}\ and\ \bibinfo {author} {\bibfnamefont {A.}~\bibnamefont
  {Marmodoro}},\ }\bibfield  {title} {\bibinfo {title} {Atomistic spin dynamics
  simulations of magnonic spin seebeck and spin nernst effects in
  altermagnets},\ }\href@noop {} {\bibfield  {journal} {\bibinfo  {journal}
  {Phys. Rev. B}\ }\textbf {\bibinfo {volume} {110}},\ \bibinfo {pages}
  {094427} (\bibinfo {year} {2024})}\BibitemShut {NoStop}%
\bibitem [{\citenamefont {Gonz{\'a}lez-Hern{\'a}ndez}\ \emph
  {et~al.}(2021)\citenamefont {Gonz{\'a}lez-Hern{\'a}ndez}, \citenamefont
  {{\v{S}}mejkal}, \citenamefont {V{\`y}born{\`y}}, \citenamefont {Yahagi},
  \citenamefont {Sinova}, \citenamefont {Jungwirth},\ and\ \citenamefont
  {{\v{Z}}elezn{\`y}}}]{gonzalez2021efficient}%
  \BibitemOpen
  \bibfield  {author} {\bibinfo {author} {\bibfnamefont {R.}~\bibnamefont
  {Gonz{\'a}lez-Hern{\'a}ndez}}, \bibinfo {author} {\bibfnamefont
  {L.}~\bibnamefont {{\v{S}}mejkal}}, \bibinfo {author} {\bibfnamefont
  {K.}~\bibnamefont {V{\`y}born{\`y}}}, \bibinfo {author} {\bibfnamefont
  {Y.}~\bibnamefont {Yahagi}}, \bibinfo {author} {\bibfnamefont
  {J.}~\bibnamefont {Sinova}}, \bibinfo {author} {\bibfnamefont
  {T.}~\bibnamefont {Jungwirth}},\ and\ \bibinfo {author} {\bibfnamefont
  {J.}~\bibnamefont {{\v{Z}}elezn{\`y}}},\ }\bibfield  {title} {\bibinfo
  {title} {Efficient electrical spin splitter based on nonrelativistic
  collinear antiferromagnetism},\ }\href@noop {} {\bibfield  {journal}
  {\bibinfo  {journal} {Phys. Rev. Lett.}\ }\textbf {\bibinfo {volume} {126}},\
  \bibinfo {pages} {127701} (\bibinfo {year} {2021})}\BibitemShut {NoStop}%
\bibitem [{\citenamefont {Bai}\ \emph {et~al.}(2023)\citenamefont {Bai},
  \citenamefont {Zhang}, \citenamefont {Zhou}, \citenamefont {Chen},
  \citenamefont {Wan}, \citenamefont {Han}, \citenamefont {Zhu}, \citenamefont
  {Liang}, \citenamefont {Su}, \citenamefont {Han}, \citenamefont {Pan},\ and\
  \citenamefont {Song}}]{PhysRevLett.130.216701}%
  \BibitemOpen
  \bibfield  {author} {\bibinfo {author} {\bibfnamefont {H.}~\bibnamefont
  {Bai}}, \bibinfo {author} {\bibfnamefont {Y.~C.}\ \bibnamefont {Zhang}},
  \bibinfo {author} {\bibfnamefont {Y.~J.}\ \bibnamefont {Zhou}}, \bibinfo
  {author} {\bibfnamefont {P.}~\bibnamefont {Chen}}, \bibinfo {author}
  {\bibfnamefont {C.~H.}\ \bibnamefont {Wan}}, \bibinfo {author} {\bibfnamefont
  {L.}~\bibnamefont {Han}}, \bibinfo {author} {\bibfnamefont {W.~X.}\
  \bibnamefont {Zhu}}, \bibinfo {author} {\bibfnamefont {S.~X.}\ \bibnamefont
  {Liang}}, \bibinfo {author} {\bibfnamefont {Y.~C.}\ \bibnamefont {Su}},
  \bibinfo {author} {\bibfnamefont {X.~F.}\ \bibnamefont {Han}}, \bibinfo
  {author} {\bibfnamefont {F.}~\bibnamefont {Pan}},\ and\ \bibinfo {author}
  {\bibfnamefont {C.}~\bibnamefont {Song}},\ }\bibfield  {title} {\bibinfo
  {title} {Efficient spin-to-charge conversion via altermagnetic spin splitting
  effect in antiferromagnet {${\mathrm{RuO}}_{2}$}},\ }\href@noop {} {\bibfield
   {journal} {\bibinfo  {journal} {Phys. Rev. Lett.}\ }\textbf {\bibinfo
  {volume} {130}},\ \bibinfo {pages} {216701} (\bibinfo {year}
  {2023})}\BibitemShut {NoStop}%
\bibitem [{\citenamefont {Sun}\ \emph {et~al.}(2025)\citenamefont {Sun},
  \citenamefont {Yang},\ and\ \citenamefont {Chen}}]{sun2025spin}%
  \BibitemOpen
  \bibfield  {author} {\bibinfo {author} {\bibfnamefont {Y.~J.}\ \bibnamefont
  {Sun}}, \bibinfo {author} {\bibfnamefont {F.}~\bibnamefont {Yang}},\ and\
  \bibinfo {author} {\bibfnamefont {L.~Q.}\ \bibnamefont {Chen}},\ }\bibfield
  {title} {\bibinfo {title} {Spin relaxation and transport behavior in $d$-wave
  altermagnetic systems},\ }\href {https://doi.org/10.1103/v12v-gl4n}
  {\bibfield  {journal} {\bibinfo  {journal} {Phys. Rev. B}\ }\textbf {\bibinfo
  {volume} {112}},\ \bibinfo {pages} {024412} (\bibinfo {year}
  {2025})}\BibitemShut {NoStop}%
\bibitem [{\citenamefont {Feng}\ \emph {et~al.}(2022)\citenamefont {Feng},
  \citenamefont {Zhou}, \citenamefont {{\v{S}}mejkal}, \citenamefont {Wu},
  \citenamefont {Zhu}, \citenamefont {Guo}, \citenamefont
  {Gonz{\'a}lez-Hern{\'a}ndez}, \citenamefont {Wang}, \citenamefont {Yan},
  \citenamefont {Qin} \emph {et~al.}}]{feng2022anomalous}%
  \BibitemOpen
  \bibfield  {author} {\bibinfo {author} {\bibfnamefont {Z.}~\bibnamefont
  {Feng}}, \bibinfo {author} {\bibfnamefont {X.}~\bibnamefont {Zhou}}, \bibinfo
  {author} {\bibfnamefont {L.}~\bibnamefont {{\v{S}}mejkal}}, \bibinfo {author}
  {\bibfnamefont {L.}~\bibnamefont {Wu}}, \bibinfo {author} {\bibfnamefont
  {Z.}~\bibnamefont {Zhu}}, \bibinfo {author} {\bibfnamefont {H.}~\bibnamefont
  {Guo}}, \bibinfo {author} {\bibfnamefont {R.}~\bibnamefont
  {Gonz{\'a}lez-Hern{\'a}ndez}}, \bibinfo {author} {\bibfnamefont
  {X.}~\bibnamefont {Wang}}, \bibinfo {author} {\bibfnamefont {H.}~\bibnamefont
  {Yan}}, \bibinfo {author} {\bibfnamefont {P.}~\bibnamefont {Qin}}, \emph
  {et~al.},\ }\bibfield  {title} {\bibinfo {title} {An anomalous hall effect in
  altermagnetic ruthenium dioxide},\ }\href@noop {} {\bibfield  {journal}
  {\bibinfo  {journal} {Nat. Electron.}\ }\textbf {\bibinfo {volume} {5}},\
  \bibinfo {pages} {735} (\bibinfo {year} {2022})}\BibitemShut {NoStop}%
\bibitem [{\citenamefont {Liao}\ \emph {et~al.}(2024)\citenamefont {Liao},
  \citenamefont {Wang}, \citenamefont {Tien}, \citenamefont {Huang},\ and\
  \citenamefont {Qu}}]{liao2024separation}%
  \BibitemOpen
  \bibfield  {author} {\bibinfo {author} {\bibfnamefont {C.-T.}\ \bibnamefont
  {Liao}}, \bibinfo {author} {\bibfnamefont {Y.-C.}\ \bibnamefont {Wang}},
  \bibinfo {author} {\bibfnamefont {Y.-C.}\ \bibnamefont {Tien}}, \bibinfo
  {author} {\bibfnamefont {S.-Y.}\ \bibnamefont {Huang}},\ and\ \bibinfo
  {author} {\bibfnamefont {D.}~\bibnamefont {Qu}},\ }\bibfield  {title}
  {\bibinfo {title} {Separation of inverse altermagnetic spin-splitting effect
  from inverse spin hall effect in ruo 2},\ }\href@noop {} {\bibfield
  {journal} {\bibinfo  {journal} {Phys. Rev. Lett.}\ }\textbf {\bibinfo
  {volume} {133}},\ \bibinfo {pages} {056701} (\bibinfo {year}
  {2024})}\BibitemShut {NoStop}%
\bibitem [{\citenamefont {Sato}\ \emph {et~al.}(2024)\citenamefont {Sato},
  \citenamefont {Haddad}, \citenamefont {Fulga}, \citenamefont {Assaad},\ and\
  \citenamefont {van~den Brink}}]{sato2024altermagnetic}%
  \BibitemOpen
  \bibfield  {author} {\bibinfo {author} {\bibfnamefont {T.}~\bibnamefont
  {Sato}}, \bibinfo {author} {\bibfnamefont {S.}~\bibnamefont {Haddad}},
  \bibinfo {author} {\bibfnamefont {I.~C.}\ \bibnamefont {Fulga}}, \bibinfo
  {author} {\bibfnamefont {F.~F.}\ \bibnamefont {Assaad}},\ and\ \bibinfo
  {author} {\bibfnamefont {J.}~\bibnamefont {van~den Brink}},\ }\bibfield
  {title} {\bibinfo {title} {Altermagnetic anomalous hall effect emerging from
  electronic correlations},\ }\href@noop {} {\bibfield  {journal} {\bibinfo
  {journal} {Phys. Rev. Lett.}\ }\textbf {\bibinfo {volume} {133}},\ \bibinfo
  {pages} {086503} (\bibinfo {year} {2024})}\BibitemShut {NoStop}%
\bibitem [{\citenamefont {Ma}\ \emph {et~al.}(2021)\citenamefont {Ma},
  \citenamefont {Hu}, \citenamefont {Li}, \citenamefont {Liu}, \citenamefont
  {Yao}, \citenamefont {Jia},\ and\ \citenamefont
  {Liu}}]{ma2021multifunctional}%
  \BibitemOpen
  \bibfield  {author} {\bibinfo {author} {\bibfnamefont {H.-Y.}\ \bibnamefont
  {Ma}}, \bibinfo {author} {\bibfnamefont {M.}~\bibnamefont {Hu}}, \bibinfo
  {author} {\bibfnamefont {N.}~\bibnamefont {Li}}, \bibinfo {author}
  {\bibfnamefont {J.}~\bibnamefont {Liu}}, \bibinfo {author} {\bibfnamefont
  {W.}~\bibnamefont {Yao}}, \bibinfo {author} {\bibfnamefont {J.-F.}\
  \bibnamefont {Jia}},\ and\ \bibinfo {author} {\bibfnamefont {J.}~\bibnamefont
  {Liu}},\ }\bibfield  {title} {\bibinfo {title} {Multifunctional
  antiferromagnetic materials with giant piezomagnetism and noncollinear spin
  current},\ }\href@noop {} {\bibfield  {journal} {\bibinfo  {journal} {Nat.
  Commun.}\ }\textbf {\bibinfo {volume} {12}},\ \bibinfo {pages} {2846}
  (\bibinfo {year} {2021})}\BibitemShut {NoStop}%
\bibitem [{\citenamefont {Fu}\ \emph {et~al.}(2025)\citenamefont {Fu},
  \citenamefont {Lv}, \citenamefont {Xu}, \citenamefont {Cayao}, \citenamefont
  {Liu},\ and\ \citenamefont {Yu}}]{fu2025all}%
  \BibitemOpen
  \bibfield  {author} {\bibinfo {author} {\bibfnamefont {P.-H.}\ \bibnamefont
  {Fu}}, \bibinfo {author} {\bibfnamefont {Q.}~\bibnamefont {Lv}}, \bibinfo
  {author} {\bibfnamefont {Y.}~\bibnamefont {Xu}}, \bibinfo {author}
  {\bibfnamefont {J.}~\bibnamefont {Cayao}}, \bibinfo {author} {\bibfnamefont
  {J.-F.}\ \bibnamefont {Liu}},\ and\ \bibinfo {author} {\bibfnamefont {X.-L.}\
  \bibnamefont {Yu}},\ }\bibfield  {title} {\bibinfo {title} {All-electrically
  controlled spintronics in altermagnetic heterostructures},\ }\href@noop {}
  {\bibfield  {journal} {\bibinfo  {journal} {arXiv:2506.05504}\ } (\bibinfo
  {year} {2025})}\BibitemShut {NoStop}%
\bibitem [{\citenamefont {Fukaya}\ \emph
  {et~al.}(2025{\natexlab{a}})\citenamefont {Fukaya}, \citenamefont {Lu},
  \citenamefont {Yada}, \citenamefont {Tanaka},\ and\ \citenamefont
  {Cayao}}]{fukaya2025superconducting}%
  \BibitemOpen
  \bibfield  {author} {\bibinfo {author} {\bibfnamefont {Y.}~\bibnamefont
  {Fukaya}}, \bibinfo {author} {\bibfnamefont {B.}~\bibnamefont {Lu}}, \bibinfo
  {author} {\bibfnamefont {K.}~\bibnamefont {Yada}}, \bibinfo {author}
  {\bibfnamefont {Y.}~\bibnamefont {Tanaka}},\ and\ \bibinfo {author}
  {\bibfnamefont {J.}~\bibnamefont {Cayao}},\ }\bibfield  {title} {\bibinfo
  {title} {Superconducting phenomena in systems with unconventional magnets},\
  }\href@noop {} {\bibfield  {journal} {\bibinfo  {journal} {arXiv:2502.15400}\
  } (\bibinfo {year} {2025}{\natexlab{a}})}\BibitemShut {NoStop}%
\bibitem [{\citenamefont {Fukaya}\ \emph
  {et~al.}(2025{\natexlab{b}})\citenamefont {Fukaya}, \citenamefont {Maeda},
  \citenamefont {Yada}, \citenamefont {Cayao}, \citenamefont {Tanaka},\ and\
  \citenamefont {Lu}}]{fukaya2025josephson}%
  \BibitemOpen
  \bibfield  {author} {\bibinfo {author} {\bibfnamefont {Y.}~\bibnamefont
  {Fukaya}}, \bibinfo {author} {\bibfnamefont {K.}~\bibnamefont {Maeda}},
  \bibinfo {author} {\bibfnamefont {K.}~\bibnamefont {Yada}}, \bibinfo {author}
  {\bibfnamefont {J.}~\bibnamefont {Cayao}}, \bibinfo {author} {\bibfnamefont
  {Y.}~\bibnamefont {Tanaka}},\ and\ \bibinfo {author} {\bibfnamefont
  {B.}~\bibnamefont {Lu}},\ }\bibfield  {title} {\bibinfo {title} {Josephson
  effect and odd-frequency pairing in superconducting junctions with
  unconventional magnets},\ }\href@noop {} {\bibfield  {journal} {\bibinfo
  {journal} {Phys. Rev. B}\ }\textbf {\bibinfo {volume} {111}},\ \bibinfo
  {pages} {064502} (\bibinfo {year} {2025}{\natexlab{b}})}\BibitemShut
  {NoStop}%
\bibitem [{\citenamefont {Zhao}\ \emph {et~al.}(2025)\citenamefont {Zhao},
  \citenamefont {Fukaya}, \citenamefont {Burset}, \citenamefont {Cayao},
  \citenamefont {Tanaka},\ and\ \citenamefont {Lu}}]{PhysRevB.111.184515}%
  \BibitemOpen
  \bibfield  {author} {\bibinfo {author} {\bibfnamefont {W.}~\bibnamefont
  {Zhao}}, \bibinfo {author} {\bibfnamefont {Y.}~\bibnamefont {Fukaya}},
  \bibinfo {author} {\bibfnamefont {P.}~\bibnamefont {Burset}}, \bibinfo
  {author} {\bibfnamefont {J.}~\bibnamefont {Cayao}}, \bibinfo {author}
  {\bibfnamefont {Y.}~\bibnamefont {Tanaka}},\ and\ \bibinfo {author}
  {\bibfnamefont {B.}~\bibnamefont {Lu}},\ }\bibfield  {title} {\bibinfo
  {title} {Orientation-dependent transport in junctions formed by $d$-wave
  altermagnets and $d$-wave superconductors},\ }\href
  {https://doi.org/10.1103/PhysRevB.111.184515} {\bibfield  {journal} {\bibinfo
   {journal} {Phys. Rev. B}\ }\textbf {\bibinfo {volume} {111}},\ \bibinfo
  {pages} {184515} (\bibinfo {year} {2025})}\BibitemShut {NoStop}%
\bibitem [{\citenamefont {Ouassou}\ \emph {et~al.}(2023)\citenamefont
  {Ouassou}, \citenamefont {Brataas},\ and\ \citenamefont
  {Linder}}]{PhysRevLett.131.076003}%
  \BibitemOpen
  \bibfield  {author} {\bibinfo {author} {\bibfnamefont {J.~A.}\ \bibnamefont
  {Ouassou}}, \bibinfo {author} {\bibfnamefont {A.}~\bibnamefont {Brataas}},\
  and\ \bibinfo {author} {\bibfnamefont {J.}~\bibnamefont {Linder}},\
  }\bibfield  {title} {\bibinfo {title} {dc josephson effect in altermagnets},\
  }\href {https://doi.org/10.1103/PhysRevLett.131.076003} {\bibfield  {journal}
  {\bibinfo  {journal} {Phys. Rev. Lett.}\ }\textbf {\bibinfo {volume} {131}},\
  \bibinfo {pages} {076003} (\bibinfo {year} {2023})}\BibitemShut {NoStop}%
\bibitem [{\citenamefont {Sun}\ \emph {et~al.}(2023)\citenamefont {Sun},
  \citenamefont {Brataas},\ and\ \citenamefont {Linder}}]{PhysRevB.108.054511}%
  \BibitemOpen
  \bibfield  {author} {\bibinfo {author} {\bibfnamefont {C.}~\bibnamefont
  {Sun}}, \bibinfo {author} {\bibfnamefont {A.}~\bibnamefont {Brataas}},\ and\
  \bibinfo {author} {\bibfnamefont {J.}~\bibnamefont {Linder}},\ }\bibfield
  {title} {\bibinfo {title} {Andreev reflection in altermagnets},\ }\href
  {https://doi.org/10.1103/PhysRevB.108.054511} {\bibfield  {journal} {\bibinfo
   {journal} {Phys. Rev. B}\ }\textbf {\bibinfo {volume} {108}},\ \bibinfo
  {pages} {054511} (\bibinfo {year} {2023})}\BibitemShut {NoStop}%
\bibitem [{\citenamefont {Wu}\ \emph {et~al.}(2025)\citenamefont {Wu},
  \citenamefont {Wang},\ and\ \citenamefont {Fernandes}}]{dlpb-gfct}%
  \BibitemOpen
  \bibfield  {author} {\bibinfo {author} {\bibfnamefont {Y.-M.}\ \bibnamefont
  {Wu}}, \bibinfo {author} {\bibfnamefont {Y.}~\bibnamefont {Wang}},\ and\
  \bibinfo {author} {\bibfnamefont {R.~M.}\ \bibnamefont {Fernandes}},\
  }\bibfield  {title} {\bibinfo {title} {Intra-unit-cell singlet pairing
  mediated by altermagnetic fluctuations},\ }\href
  {https://doi.org/10.1103/dlpb-gfct} {\bibfield  {journal} {\bibinfo
  {journal} {Phys. Rev. Lett.}\ }\textbf {\bibinfo {volume} {135}},\ \bibinfo
  {pages} {156001} (\bibinfo {year} {2025})}\BibitemShut {NoStop}%
\bibitem [{\citenamefont {Zhang}\ \emph {et~al.}(2024)\citenamefont {Zhang},
  \citenamefont {Hu},\ and\ \citenamefont {Neupert}}]{zhang2024finite}%
  \BibitemOpen
  \bibfield  {author} {\bibinfo {author} {\bibfnamefont {S.-B.}\ \bibnamefont
  {Zhang}}, \bibinfo {author} {\bibfnamefont {L.-H.}\ \bibnamefont {Hu}},\ and\
  \bibinfo {author} {\bibfnamefont {T.}~\bibnamefont {Neupert}},\ }\bibfield
  {title} {\bibinfo {title} {Finite-momentum cooper pairing in proximitized
  altermagnets},\ }\href@noop {} {\bibfield  {journal} {\bibinfo  {journal}
  {Nat. Commun.}\ }\textbf {\bibinfo {volume} {15}},\ \bibinfo {pages} {1801}
  (\bibinfo {year} {2024})}\BibitemShut {NoStop}%
\bibitem [{\citenamefont {Hu}\ \emph {et~al.}(2025)\citenamefont {Hu},
  \citenamefont {Liu},\ and\ \citenamefont {Liu}}]{hu2025unconventional}%
  \BibitemOpen
  \bibfield  {author} {\bibinfo {author} {\bibfnamefont {H.}~\bibnamefont
  {Hu}}, \bibinfo {author} {\bibfnamefont {Z.}~\bibnamefont {Liu}},\ and\
  \bibinfo {author} {\bibfnamefont {X.-J.}\ \bibnamefont {Liu}},\ }\bibfield
  {title} {\bibinfo {title} {Unconventional superconductivity of an
  altermagnetic metal: Polarized bcs and inhomogeneous
  fulde-ferrell-larkin-ovchinnikov states},\ }\href@noop {} {\bibfield
  {journal} {\bibinfo  {journal} {arXiv:2505.10196}\ } (\bibinfo {year}
  {2025})}\BibitemShut {NoStop}%
\bibitem [{\citenamefont {Iorsh}(2025)}]{6wxh-p4mc}%
  \BibitemOpen
  \bibfield  {author} {\bibinfo {author} {\bibfnamefont {I.~V.}\ \bibnamefont
  {Iorsh}},\ }\bibfield  {title} {\bibinfo {title} {Electron pairing by
  dispersive phonons in altermagnets: Reentrant superconductivity and
  continuous transition to finite momentum superconducting state},\ }\href
  {https://doi.org/10.1103/6wxh-p4mc} {\bibfield  {journal} {\bibinfo
  {journal} {Phys. Rev. B}\ }\textbf {\bibinfo {volume} {111}},\ \bibinfo
  {pages} {L220503} (\bibinfo {year} {2025})}\BibitemShut {NoStop}%
\bibitem [{\citenamefont {Mukasa}\ and\ \citenamefont
  {Masaki}(2025)}]{mukasa2025finite}%
  \BibitemOpen
  \bibfield  {author} {\bibinfo {author} {\bibfnamefont {K.}~\bibnamefont
  {Mukasa}}\ and\ \bibinfo {author} {\bibfnamefont {Y.}~\bibnamefont
  {Masaki}},\ }\bibfield  {title} {\bibinfo {title} {Finite-momentum
  superconductivity in two-dimensional altermagnets with a rashba-type
  spin--orbit coupling},\ }\href@noop {} {\bibfield  {journal} {\bibinfo
  {journal} {J. Phys. Soc. Jpn.}\ }\textbf {\bibinfo {volume} {94}},\ \bibinfo
  {pages} {064705} (\bibinfo {year} {2025})}\BibitemShut {NoStop}%
\bibitem [{\citenamefont {Sim}\ and\ \citenamefont
  {Knolle}(2024)}]{sim2024pair}%
  \BibitemOpen
  \bibfield  {author} {\bibinfo {author} {\bibfnamefont {G.}~\bibnamefont
  {Sim}}\ and\ \bibinfo {author} {\bibfnamefont {J.}~\bibnamefont {Knolle}},\
  }\bibfield  {title} {\bibinfo {title} {Pair density waves and supercurrent
  diode effect in altermagnets},\ }\href@noop {} {\bibfield  {journal}
  {\bibinfo  {journal} {arXiv:2407.01513}\ } (\bibinfo {year}
  {2024})}\BibitemShut {NoStop}%
\bibitem [{\citenamefont {Sumita}\ \emph {et~al.}(2023)\citenamefont {Sumita},
  \citenamefont {Naka},\ and\ \citenamefont {Seo}}]{PhysRevResearch.5.043171}%
  \BibitemOpen
  \bibfield  {author} {\bibinfo {author} {\bibfnamefont {S.}~\bibnamefont
  {Sumita}}, \bibinfo {author} {\bibfnamefont {M.}~\bibnamefont {Naka}},\ and\
  \bibinfo {author} {\bibfnamefont {H.}~\bibnamefont {Seo}},\ }\bibfield
  {title} {\bibinfo {title} {Fulde-ferrell-larkin-ovchinnikov state induced by
  antiferromagnetic order in $\ensuremath{\kappa}$-type organic conductors},\
  }\href {https://doi.org/10.1103/PhysRevResearch.5.043171} {\bibfield
  {journal} {\bibinfo  {journal} {Phys. Rev. Res.}\ }\textbf {\bibinfo {volume}
  {5}},\ \bibinfo {pages} {043171} (\bibinfo {year} {2023})}\BibitemShut
  {NoStop}%
\bibitem [{\citenamefont {Hu}\ and\ \citenamefont {Liu}(2025)}]{hu2025quantum}%
  \BibitemOpen
  \bibfield  {author} {\bibinfo {author} {\bibfnamefont {H.}~\bibnamefont
  {Hu}}\ and\ \bibinfo {author} {\bibfnamefont {X.-J.}\ \bibnamefont {Liu}},\
  }\bibfield  {title} {\bibinfo {title} {Quantum lifshitz points in an
  altermagnetic metal},\ }\href@noop {} {\bibfield  {journal} {\bibinfo
  {journal} {arXiv:2505.10242}\ } (\bibinfo {year} {2025})}\BibitemShut
  {NoStop}%
\bibitem [{\citenamefont {Yang}\ and\ \citenamefont
  {Wu}(2017)}]{PhysRevB.95.075304}%
  \BibitemOpen
  \bibfield  {author} {\bibinfo {author} {\bibfnamefont {F.}~\bibnamefont
  {Yang}}\ and\ \bibinfo {author} {\bibfnamefont {M.~W.}\ \bibnamefont {Wu}},\
  }\bibfield  {title} {\bibinfo {title} {Gapped superconductivity with all
  symmetries in insb (110) quantum wells in proximity to $s$-wave
  superconductor in fulde-ferrell-larkin-ovchinnikov phase or with a
  supercurrent},\ }\href {https://doi.org/10.1103/PhysRevB.95.075304}
  {\bibfield  {journal} {\bibinfo  {journal} {Phys. Rev. B}\ }\textbf {\bibinfo
  {volume} {95}},\ \bibinfo {pages} {075304} (\bibinfo {year}
  {2017})}\BibitemShut {NoStop}%
\bibitem [{\citenamefont {Stanescu}\ \emph {et~al.}(2011)\citenamefont
  {Stanescu}, \citenamefont {Lutchyn},\ and\ \citenamefont
  {Das~Sarma}}]{PhysRevB.84.144522}%
  \BibitemOpen
  \bibfield  {author} {\bibinfo {author} {\bibfnamefont {T.~D.}\ \bibnamefont
  {Stanescu}}, \bibinfo {author} {\bibfnamefont {R.~M.}\ \bibnamefont
  {Lutchyn}},\ and\ \bibinfo {author} {\bibfnamefont {S.}~\bibnamefont
  {Das~Sarma}},\ }\bibfield  {title} {\bibinfo {title} {Majorana fermions in
  semiconductor nanowires},\ }\href
  {https://doi.org/10.1103/PhysRevB.84.144522} {\bibfield  {journal} {\bibinfo
  {journal} {Phys. Rev. B}\ }\textbf {\bibinfo {volume} {84}},\ \bibinfo
  {pages} {144522} (\bibinfo {year} {2011})}\BibitemShut {NoStop}%
\bibitem [{\citenamefont {Danon}\ and\ \citenamefont
  {Flensberg}(2015)}]{PhysRevB.91.165425}%
  \BibitemOpen
  \bibfield  {author} {\bibinfo {author} {\bibfnamefont {J.}~\bibnamefont
  {Danon}}\ and\ \bibinfo {author} {\bibfnamefont {K.}~\bibnamefont
  {Flensberg}},\ }\bibfield  {title} {\bibinfo {title} {Interaction effects on
  proximity-induced superconductivity in semiconducting nanowires},\ }\href
  {https://doi.org/10.1103/PhysRevB.91.165425} {\bibfield  {journal} {\bibinfo
  {journal} {Phys. Rev. B}\ }\textbf {\bibinfo {volume} {91}},\ \bibinfo
  {pages} {165425} (\bibinfo {year} {2015})}\BibitemShut {NoStop}%
\bibitem [{\citenamefont {Island}\ \emph {et~al.}(2017)\citenamefont {Island},
  \citenamefont {Gaudenzi}, \citenamefont {de~Bruijckere}, \citenamefont
  {Burzur\'{\i}}, \citenamefont {Franco}, \citenamefont {Mas-Torrent},
  \citenamefont {Rovira}, \citenamefont {Veciana}, \citenamefont {Klapwijk},
  \citenamefont {Aguado},\ and\ \citenamefont {van~der
  Zant}}]{PhysRevLett.118.117001}%
  \BibitemOpen
  \bibfield  {author} {\bibinfo {author} {\bibfnamefont {J.~O.}\ \bibnamefont
  {Island}}, \bibinfo {author} {\bibfnamefont {R.}~\bibnamefont {Gaudenzi}},
  \bibinfo {author} {\bibfnamefont {J.}~\bibnamefont {de~Bruijckere}}, \bibinfo
  {author} {\bibfnamefont {E.}~\bibnamefont {Burzur\'{\i}}}, \bibinfo {author}
  {\bibfnamefont {C.}~\bibnamefont {Franco}}, \bibinfo {author} {\bibfnamefont
  {M.}~\bibnamefont {Mas-Torrent}}, \bibinfo {author} {\bibfnamefont
  {C.}~\bibnamefont {Rovira}}, \bibinfo {author} {\bibfnamefont
  {J.}~\bibnamefont {Veciana}}, \bibinfo {author} {\bibfnamefont {T.~M.}\
  \bibnamefont {Klapwijk}}, \bibinfo {author} {\bibfnamefont {R.}~\bibnamefont
  {Aguado}},\ and\ \bibinfo {author} {\bibfnamefont {H.~S.~J.}\ \bibnamefont
  {van~der Zant}},\ }\bibfield  {title} {\bibinfo {title} {Proximity-induced
  shiba states in a molecular junction},\ }\href
  {https://doi.org/10.1103/PhysRevLett.118.117001} {\bibfield  {journal}
  {\bibinfo  {journal} {Phys. Rev. Lett.}\ }\textbf {\bibinfo {volume} {118}},\
  \bibinfo {pages} {117001} (\bibinfo {year} {2017})}\BibitemShut {NoStop}%
\bibitem [{\citenamefont {Buzdin}(2005)}]{RevModPhys.77.935}%
  \BibitemOpen
  \bibfield  {author} {\bibinfo {author} {\bibfnamefont {A.~I.}\ \bibnamefont
  {Buzdin}},\ }\bibfield  {title} {\bibinfo {title} {Proximity effects in
  superconductor-ferromagnet heterostructures},\ }\href
  {https://doi.org/10.1103/RevModPhys.77.935} {\bibfield  {journal} {\bibinfo
  {journal} {Rev. Mod. Phys.}\ }\textbf {\bibinfo {volume} {77}},\ \bibinfo
  {pages} {935} (\bibinfo {year} {2005})}\BibitemShut {NoStop}%
\bibitem [{\citenamefont {Noda}\ \emph {et~al.}(2016)\citenamefont {Noda},
  \citenamefont {Ohno},\ and\ \citenamefont {Nakamura}}]{noda2016momentum}%
  \BibitemOpen
  \bibfield  {author} {\bibinfo {author} {\bibfnamefont {Y.}~\bibnamefont
  {Noda}}, \bibinfo {author} {\bibfnamefont {K.}~\bibnamefont {Ohno}},\ and\
  \bibinfo {author} {\bibfnamefont {S.}~\bibnamefont {Nakamura}},\ }\bibfield
  {title} {\bibinfo {title} {Momentum-dependent band spin splitting in
  semiconducting {${\mathrm{MnO}}_{2}$}: a density functional calculation},\
  }\href@noop {} {\bibfield  {journal} {\bibinfo  {journal} {Phys. Chem. Chem.
  Phys.}\ }\textbf {\bibinfo {volume} {18}},\ \bibinfo {pages} {13294}
  (\bibinfo {year} {2016})}\BibitemShut {NoStop}%
\bibitem [{\citenamefont {Schrieffer}(1964)}]{schrieffer1964theory}%
  \BibitemOpen
  \bibfield  {author} {\bibinfo {author} {\bibfnamefont {J.}~\bibnamefont
  {Schrieffer}},\ }\href@noop {} {\emph {\bibinfo {title} {Theory of
  Superconductivity}}}\ (\bibinfo  {publisher} {W.A. Benjamin},\ \bibinfo
  {year} {1964})\BibitemShut {NoStop}%
\bibitem [{\citenamefont {Bardeen}\ \emph {et~al.}(1957)\citenamefont
  {Bardeen}, \citenamefont {Cooper},\ and\ \citenamefont
  {Schrieffer}}]{bardeen1957theory}%
  \BibitemOpen
  \bibfield  {author} {\bibinfo {author} {\bibfnamefont {J.}~\bibnamefont
  {Bardeen}}, \bibinfo {author} {\bibfnamefont {L.~N.}\ \bibnamefont
  {Cooper}},\ and\ \bibinfo {author} {\bibfnamefont {J.~R.}\ \bibnamefont
  {Schrieffer}},\ }\bibfield  {title} {\bibinfo {title} {Theory of
  superconductivity},\ }\href@noop {} {\bibfield  {journal} {\bibinfo
  {journal} {Phys. Rev.}\ }\textbf {\bibinfo {volume} {108}},\ \bibinfo {pages}
  {1175} (\bibinfo {year} {1957})}\BibitemShut {NoStop}%
\bibitem [{\citenamefont {Li}\ \emph {et~al.}(2021)\citenamefont {Li},
  \citenamefont {Kivelson},\ and\ \citenamefont {Lee}}]{li21superconductor}%
  \BibitemOpen
  \bibfield  {author} {\bibinfo {author} {\bibfnamefont {Z.-X.}\ \bibnamefont
  {Li}}, \bibinfo {author} {\bibfnamefont {S.~A.}\ \bibnamefont {Kivelson}},\
  and\ \bibinfo {author} {\bibfnamefont {D.-H.}\ \bibnamefont {Lee}},\
  }\bibfield  {title} {\bibinfo {title} {Superconductor-to-metal transition in
  overdoped cuprates},\ }\href {https://doi.org/10.1038/s41535-021-00335-4}
  {\bibfield  {journal} {\bibinfo  {journal} {npj Quantum Materials}\ }\textbf
  {\bibinfo {volume} {6}},\ \bibinfo {pages} {36} (\bibinfo {year}
  {2021})}\BibitemShut {NoStop}%
\bibitem [{\citenamefont {Yang}\ and\ \citenamefont
  {Chen}(2024)}]{yang24thermodynamic}%
  \BibitemOpen
  \bibfield  {author} {\bibinfo {author} {\bibfnamefont {F.}~\bibnamefont
  {Yang}}\ and\ \bibinfo {author} {\bibfnamefont {L.~Q.}\ \bibnamefont
  {Chen}},\ }\href {https://arxiv.org/abs/2410.05216} {\bibinfo {title}
  {Thermodynamic theory of disordered 2d superconductors}} (\bibinfo {year}
  {2024}),\ \Eprint {https://arxiv.org/abs/2410.05216} {arXiv:2410.05216}
  \BibitemShut {NoStop}%
\bibitem [{\citenamefont {Kosztin}\ \emph {et~al.}(1998)\citenamefont
  {Kosztin}, \citenamefont {Kos}, \citenamefont {Stone},\ and\ \citenamefont
  {Leggett}}]{PhysRevB.58.9365}%
  \BibitemOpen
  \bibfield  {author} {\bibinfo {author} {\bibfnamefont {I.}~\bibnamefont
  {Kosztin}}, \bibinfo {author} {\bibfnamefont {i.~c.~v.}\ \bibnamefont {Kos}},
  \bibinfo {author} {\bibfnamefont {M.}~\bibnamefont {Stone}},\ and\ \bibinfo
  {author} {\bibfnamefont {A.~J.}\ \bibnamefont {Leggett}},\ }\bibfield
  {title} {\bibinfo {title} {Free energy of an inhomogeneous superconductor: A
  wave-function approach},\ }\href {https://doi.org/10.1103/PhysRevB.58.9365}
  {\bibfield  {journal} {\bibinfo  {journal} {Phys. Rev. B}\ }\textbf {\bibinfo
  {volume} {58}},\ \bibinfo {pages} {9365} (\bibinfo {year}
  {1998})}\BibitemShut {NoStop}%
\bibitem [{\citenamefont {Jafari}(2024)}]{PhysRevResearch.6.033006}%
  \BibitemOpen
  \bibfield  {author} {\bibinfo {author} {\bibfnamefont {S.~A.}\ \bibnamefont
  {Jafari}},\ }\bibfield  {title} {\bibinfo {title} {Moving frame theory of
  zero-bias photocurrent on the surface of topological insulators},\ }\href
  {https://doi.org/10.1103/PhysRevResearch.6.033006} {\bibfield  {journal}
  {\bibinfo  {journal} {Phys. Rev. Res.}\ }\textbf {\bibinfo {volume} {6}},\
  \bibinfo {pages} {033006} (\bibinfo {year} {2024})}\BibitemShut {NoStop}%
\bibitem [{\citenamefont {Yang}\ and\ \citenamefont {Wu}(2021)}]{yang21theory}%
  \BibitemOpen
  \bibfield  {author} {\bibinfo {author} {\bibfnamefont {F.}~\bibnamefont
  {Yang}}\ and\ \bibinfo {author} {\bibfnamefont {M.~W.}\ \bibnamefont {Wu}},\
  }\bibfield  {title} {\bibinfo {title} {Theory of coupled dual dynamics of
  macroscopic phase coherence and microscopic electronic fluids: Effect of
  dephasing on cuprate superconductivity},\ }\href
  {https://doi.org/10.1103/PhysRevB.104.214510} {\bibfield  {journal} {\bibinfo
   {journal} {Phys. Rev. B}\ }\textbf {\bibinfo {volume} {104}},\ \bibinfo
  {pages} {214510} (\bibinfo {year} {2021})}\BibitemShut {NoStop}%
\bibitem [{\citenamefont {Yang}\ and\ \citenamefont
  {Wu}(2018{\natexlab{b}})}]{yang2018gauge}%
  \BibitemOpen
  \bibfield  {author} {\bibinfo {author} {\bibfnamefont {F.}~\bibnamefont
  {Yang}}\ and\ \bibinfo {author} {\bibfnamefont {M.}~\bibnamefont {Wu}},\
  }\bibfield  {title} {\bibinfo {title} {Gauge-invariant microscopic kinetic
  theory of superconductivity in response to electromagnetic fields},\
  }\href@noop {} {\bibfield  {journal} {\bibinfo  {journal} {Phys. Rev. B}\
  }\textbf {\bibinfo {volume} {98}},\ \bibinfo {pages} {094507} (\bibinfo
  {year} {2018}{\natexlab{b}})}\BibitemShut {NoStop}%
\bibitem [{\citenamefont {Littlewood}\ and\ \citenamefont
  {Varma}(1981{\natexlab{a}})}]{littlewood1981gauge}%
  \BibitemOpen
  \bibfield  {author} {\bibinfo {author} {\bibfnamefont {P.}~\bibnamefont
  {Littlewood}}\ and\ \bibinfo {author} {\bibfnamefont {C.}~\bibnamefont
  {Varma}},\ }\bibfield  {title} {\bibinfo {title} {Gauge-invariant theory of
  the dynamical interaction of charge density waves and superconductivity},\
  }\href@noop {} {\bibfield  {journal} {\bibinfo  {journal} {Phys. Rev. Lett.}\
  }\textbf {\bibinfo {volume} {47}},\ \bibinfo {pages} {811} (\bibinfo {year}
  {1981}{\natexlab{a}})}\BibitemShut {NoStop}%
\bibitem [{\citenamefont {Yang}\ and\ \citenamefont {Wu}(2019)}]{yang19gauge}%
  \BibitemOpen
  \bibfield  {author} {\bibinfo {author} {\bibfnamefont {F.}~\bibnamefont
  {Yang}}\ and\ \bibinfo {author} {\bibfnamefont {M.~W.}\ \bibnamefont {Wu}},\
  }\bibfield  {title} {\bibinfo {title} {Gauge-invariant microscopic kinetic
  theory of superconductivity: Application to the optical response of
  nambu-goldstone and higgs modes},\ }\href
  {https://doi.org/10.1103/PhysRevB.100.104513} {\bibfield  {journal} {\bibinfo
   {journal} {Phys. Rev. B}\ }\textbf {\bibinfo {volume} {100}},\ \bibinfo
  {pages} {104513} (\bibinfo {year} {2019})}\BibitemShut {NoStop}%
\bibitem [{\citenamefont {Nambu}(1960)}]{nambu1960quasi}%
  \BibitemOpen
  \bibfield  {author} {\bibinfo {author} {\bibfnamefont {Y.}~\bibnamefont
  {Nambu}},\ }\bibfield  {title} {\bibinfo {title} {Quasi-particles and gauge
  invariance in the theory of superconductivity},\ }\href
  {https://journals.aps.org/pr/abstract/10.1103/PhysRev.117.648} {\bibfield
  {journal} {\bibinfo  {journal} {Phys. Rev.}\ }\textbf {\bibinfo {volume}
  {117}},\ \bibinfo {pages} {648} (\bibinfo {year} {1960})}\BibitemShut
  {NoStop}%
\bibitem [{\citenamefont {Sun}\ \emph {et~al.}(2020)\citenamefont {Sun},
  \citenamefont {Fogler}, \citenamefont {Basov},\ and\ \citenamefont
  {Millis}}]{sun20collective}%
  \BibitemOpen
  \bibfield  {author} {\bibinfo {author} {\bibfnamefont {Z.}~\bibnamefont
  {Sun}}, \bibinfo {author} {\bibfnamefont {M.~M.}\ \bibnamefont {Fogler}},
  \bibinfo {author} {\bibfnamefont {D.~N.}\ \bibnamefont {Basov}},\ and\
  \bibinfo {author} {\bibfnamefont {A.~J.}\ \bibnamefont {Millis}},\ }\bibfield
   {title} {\bibinfo {title} {Collective modes and terahertz near-field
  response of superconductors},\ }\href
  {https://doi.org/10.1103/PhysRevResearch.2.023413} {\bibfield  {journal}
  {\bibinfo  {journal} {Phys. Rev. Res.}\ }\textbf {\bibinfo {volume} {2}},\
  \bibinfo {pages} {023413} (\bibinfo {year} {2020})}\BibitemShut {NoStop}%
\bibitem [{\citenamefont {Littlewood}\ and\ \citenamefont
  {Varma}(1981{\natexlab{b}})}]{littlewood81gauge}%
  \BibitemOpen
  \bibfield  {author} {\bibinfo {author} {\bibfnamefont {P.~B.}\ \bibnamefont
  {Littlewood}}\ and\ \bibinfo {author} {\bibfnamefont {C.~M.}\ \bibnamefont
  {Varma}},\ }\bibfield  {title} {\bibinfo {title} {Gauge-invariant theory of
  the dynamical interaction of charge density waves and superconductivity},\
  }\href {https://doi.org/10.1103/PhysRevLett.47.811} {\bibfield  {journal}
  {\bibinfo  {journal} {Phys. Rev. Lett.}\ }\textbf {\bibinfo {volume} {47}},\
  \bibinfo {pages} {811} (\bibinfo {year} {1981}{\natexlab{b}})}\BibitemShut
  {NoStop}%
\bibitem [{\citenamefont {Ambegaokar}\ and\ \citenamefont
  {Kadanoff}(1961)}]{ambegaokar61electromagnetic}%
  \BibitemOpen
  \bibfield  {author} {\bibinfo {author} {\bibfnamefont {V.}~\bibnamefont
  {Ambegaokar}}\ and\ \bibinfo {author} {\bibfnamefont {L.~P.}\ \bibnamefont
  {Kadanoff}},\ }\bibfield  {title} {\bibinfo {title} {Electromagnetic
  properties of superconductors},\ }\href {https://doi.org/10.1007/BF02787879}
  {\bibfield  {journal} {\bibinfo  {journal} {Il Nuovo Cimento}\ }\textbf
  {\bibinfo {volume} {22}},\ \bibinfo {pages} {914} (\bibinfo {year}
  {1961})}\BibitemShut {NoStop}%
\bibitem [{\citenamefont {Yang}\ and\ \citenamefont
  {Wu}(2020)}]{PhysRevB.102.144508}%
  \BibitemOpen
  \bibfield  {author} {\bibinfo {author} {\bibfnamefont {F.}~\bibnamefont
  {Yang}}\ and\ \bibinfo {author} {\bibfnamefont {M.~W.}\ \bibnamefont {Wu}},\
  }\bibfield  {title} {\bibinfo {title} {Influence of scattering on the optical
  response of superconductors},\ }\href
  {https://doi.org/10.1103/PhysRevB.102.144508} {\bibfield  {journal} {\bibinfo
   {journal} {Phys. Rev. B}\ }\textbf {\bibinfo {volume} {102}},\ \bibinfo
  {pages} {144508} (\bibinfo {year} {2020})}\BibitemShut {NoStop}%
\bibitem [{SDC()}]{SDC}%
  \BibitemOpen
  \href@noop {} {}\bibinfo {note} {The expression of the superfluid density has
  been obtained using various theoretical approaches (e.g., diagrammatic
  formulation~\cite{PhysRevB.106.144509,PhysRevB.109.064508}, gauge-invariant
  kinetic equation
  method~\cite{yang19gauge,yang2018gauge,yang21theory,PhysRevB.102.144508},
  path-integral
  approach~\cite{sun20collective,yang24thermodynamic,yang21theory}). The
  specific expression of the superfluid density in the present context is
  presented in the Supplemental Materials}\BibitemShut {NoStop}%
\bibitem [{\citenamefont {Yang}\ and\ \citenamefont
  {Wu}(2024{\natexlab{a}})}]{PhysRevB.109.064508}%
  \BibitemOpen
  \bibfield  {author} {\bibinfo {author} {\bibfnamefont {F.}~\bibnamefont
  {Yang}}\ and\ \bibinfo {author} {\bibfnamefont {M.~W.}\ \bibnamefont {Wu}},\
  }\bibfield  {title} {\bibinfo {title} {Diamagnetic property and optical
  absorption of conventional superconductors with magnetic impurities in linear
  response},\ }\href {https://doi.org/10.1103/PhysRevB.109.064508} {\bibfield
  {journal} {\bibinfo  {journal} {Phys. Rev. B}\ }\textbf {\bibinfo {volume}
  {109}},\ \bibinfo {pages} {064508} (\bibinfo {year}
  {2024}{\natexlab{a}})}\BibitemShut {NoStop}%
\bibitem [{\citenamefont {Yang}\ and\ \citenamefont
  {Wu}(2022)}]{PhysRevB.106.144509}%
  \BibitemOpen
  \bibfield  {author} {\bibinfo {author} {\bibfnamefont {F.}~\bibnamefont
  {Yang}}\ and\ \bibinfo {author} {\bibfnamefont {M.~W.}\ \bibnamefont {Wu}},\
  }\bibfield  {title} {\bibinfo {title} {Impurity scattering in superconductors
  revisited: Diagrammatic formulation of the supercurrent-supercurrent
  correlation and higgs-mode damping},\ }\href
  {https://doi.org/10.1103/PhysRevB.106.144509} {\bibfield  {journal} {\bibinfo
   {journal} {Phys. Rev. B}\ }\textbf {\bibinfo {volume} {106}},\ \bibinfo
  {pages} {144509} (\bibinfo {year} {2022})}\BibitemShut {NoStop}%
\bibitem [{\citenamefont {Chockalingam}\ \emph {et~al.}(2008)\citenamefont
  {Chockalingam}, \citenamefont {Chand}, \citenamefont {Jesudasan},
  \citenamefont {Tripathi},\ and\ \citenamefont
  {Raychaudhuri}}]{chockalingam2008superconducting}%
  \BibitemOpen
  \bibfield  {author} {\bibinfo {author} {\bibfnamefont {S.}~\bibnamefont
  {Chockalingam}}, \bibinfo {author} {\bibfnamefont {M.}~\bibnamefont {Chand}},
  \bibinfo {author} {\bibfnamefont {J.}~\bibnamefont {Jesudasan}}, \bibinfo
  {author} {\bibfnamefont {V.}~\bibnamefont {Tripathi}},\ and\ \bibinfo
  {author} {\bibfnamefont {P.}~\bibnamefont {Raychaudhuri}},\ }\bibfield
  {title} {\bibinfo {title} {Superconducting properties and hall effect of
  epitaxial {NbN thin films}},\ }\href@noop {} {\bibfield  {journal} {\bibinfo
  {journal} {Phys. Rev. B}\ }\textbf {\bibinfo {volume} {77}},\ \bibinfo
  {pages} {214503} (\bibinfo {year} {2008})}\BibitemShut {NoStop}%
\bibitem [{\citenamefont {Matsunaga}\ \emph {et~al.}(2014)\citenamefont
  {Matsunaga}, \citenamefont {Tsuji}, \citenamefont {Fujita}, \citenamefont
  {Sugioka}, \citenamefont {Makise}, \citenamefont {Uzawa}, \citenamefont
  {Terai}, \citenamefont {Wang}, \citenamefont {Aoki},\ and\ \citenamefont
  {Shimano}}]{matsunaga2014light}%
  \BibitemOpen
  \bibfield  {author} {\bibinfo {author} {\bibfnamefont {R.}~\bibnamefont
  {Matsunaga}}, \bibinfo {author} {\bibfnamefont {N.}~\bibnamefont {Tsuji}},
  \bibinfo {author} {\bibfnamefont {H.}~\bibnamefont {Fujita}}, \bibinfo
  {author} {\bibfnamefont {A.}~\bibnamefont {Sugioka}}, \bibinfo {author}
  {\bibfnamefont {K.}~\bibnamefont {Makise}}, \bibinfo {author} {\bibfnamefont
  {Y.}~\bibnamefont {Uzawa}}, \bibinfo {author} {\bibfnamefont
  {H.}~\bibnamefont {Terai}}, \bibinfo {author} {\bibfnamefont
  {Z.}~\bibnamefont {Wang}}, \bibinfo {author} {\bibfnamefont {H.}~\bibnamefont
  {Aoki}},\ and\ \bibinfo {author} {\bibfnamefont {R.}~\bibnamefont
  {Shimano}},\ }\bibfield  {title} {\bibinfo {title} {Light-induced collective
  pseudospin precession resonating with {Higgs} mode in a superconductor},\
  }\href@noop {} {\bibfield  {journal} {\bibinfo  {journal} {Science}\ }\textbf
  {\bibinfo {volume} {345}},\ \bibinfo {pages} {1145} (\bibinfo {year}
  {2014})}\BibitemShut {NoStop}%
\bibitem [{\citenamefont {Mahan}(2013)}]{mahan2013many}%
  \BibitemOpen
  \bibfield  {author} {\bibinfo {author} {\bibfnamefont {G.~D.}\ \bibnamefont
  {Mahan}},\ }\href@noop {} {\emph {\bibinfo {title} {Many-particle physics}}}\
  (\bibinfo  {publisher} {Springer Science \& Business Media},\ \bibinfo {year}
  {2013})\BibitemShut {NoStop}%
\bibitem [{\citenamefont {Landau}\ \emph {et~al.}(1980)\citenamefont {Landau},
  \citenamefont {Lifshitz},\ and\ \citenamefont {Pitaevskii}}]{Landaubook}%
  \BibitemOpen
  \bibfield  {author} {\bibinfo {author} {\bibfnamefont {L.~D.}\ \bibnamefont
  {Landau}}, \bibinfo {author} {\bibfnamefont {E.~M.}\ \bibnamefont
  {Lifshitz}},\ and\ \bibinfo {author} {\bibfnamefont {L.~P.}\ \bibnamefont
  {Pitaevskii}},\ }\href@noop {} {\emph {\bibinfo {title} {Statistical Physics,
  Part 1}}}\ (\bibinfo  {publisher} {Pergamon, New York},\ \bibinfo {year}
  {1980})\BibitemShut {NoStop}%
\bibitem [{\citenamefont {Yang}\ and\ \citenamefont
  {Wu}(2023)}]{yang2023optical}%
  \BibitemOpen
  \bibfield  {author} {\bibinfo {author} {\bibfnamefont {F.}~\bibnamefont
  {Yang}}\ and\ \bibinfo {author} {\bibfnamefont {M.}~\bibnamefont {Wu}},\
  }\bibfield  {title} {\bibinfo {title} {Optical response of higgs mode in
  superconductors at clean limit},\ }\href@noop {} {\bibfield  {journal}
  {\bibinfo  {journal} {Ann. Phys.}\ }\textbf {\bibinfo {volume} {453}},\
  \bibinfo {pages} {169312} (\bibinfo {year} {2023})}\BibitemShut {NoStop}%
\bibitem [{\citenamefont {Yang}\ and\ \citenamefont
  {Wu}(2024{\natexlab{b}})}]{yang2024optical}%
  \BibitemOpen
  \bibfield  {author} {\bibinfo {author} {\bibfnamefont {F.}~\bibnamefont
  {Yang}}\ and\ \bibinfo {author} {\bibfnamefont {M.}~\bibnamefont {Wu}},\
  }\bibfield  {title} {\bibinfo {title} {Optical response of higgs mode in
  superconductors at clean limit: formulation through eilenberger equation and
  ginzburg--landau lagrangian},\ }\href@noop {} {\bibfield  {journal} {\bibinfo
   {journal} {J. Phys.: Condens. Matter.}\ }\textbf {\bibinfo {volume} {36}},\
  \bibinfo {pages} {425701} (\bibinfo {year} {2024}{\natexlab{b}})}\BibitemShut
  {NoStop}%
\bibitem [{sup()}]{supple}%
  \BibitemOpen
  \href@noop {} {}\bibinfo {note} {See Supplemental Material for the details of
  numerical simulation and the discussion of spin-triplet state}\BibitemShut
  {NoStop}%
\bibitem [{\citenamefont {Pekker}\ and\ \citenamefont
  {Varma}(2015)}]{pekker2015amplitude}%
  \BibitemOpen
  \bibfield  {author} {\bibinfo {author} {\bibfnamefont {D.}~\bibnamefont
  {Pekker}}\ and\ \bibinfo {author} {\bibfnamefont {C.}~\bibnamefont {Varma}},\
  }\bibfield  {title} {\bibinfo {title} {Amplitude/{Higgs} modes in condensed
  matter physics},\ }\href@noop {} {\bibfield  {journal} {\bibinfo  {journal}
  {Annu. Rev. Condens. Matter Phys.}\ }\textbf {\bibinfo {volume} {6}},\
  \bibinfo {pages} {269} (\bibinfo {year} {2015})}\BibitemShut {NoStop}%
\bibitem [{\citenamefont {Jabusch}\ \emph {et~al.}(2025)\citenamefont
  {Jabusch}, \citenamefont {Kokkinis},\ and\ \citenamefont
  {Chubukov}}]{PhysRevB.111.174507}%
  \BibitemOpen
  \bibfield  {author} {\bibinfo {author} {\bibfnamefont {N.~J.}\ \bibnamefont
  {Jabusch}}, \bibinfo {author} {\bibfnamefont {E.~K.}\ \bibnamefont
  {Kokkinis}},\ and\ \bibinfo {author} {\bibfnamefont {A.~V.}\ \bibnamefont
  {Chubukov}},\ }\bibfield  {title} {\bibinfo {title} {First order phase
  transition in a two-dimensional superconductor},\ }\href
  {https://doi.org/10.1103/PhysRevB.111.174507} {\bibfield  {journal} {\bibinfo
   {journal} {Phys. Rev. B}\ }\textbf {\bibinfo {volume} {111}},\ \bibinfo
  {pages} {174507} (\bibinfo {year} {2025})}\BibitemShut {NoStop}%
\bibitem [{\citenamefont {Faraei}\ and\ \citenamefont
  {Jafari}(2017)}]{PhysRevB.96.134516}%
  \BibitemOpen
  \bibfield  {author} {\bibinfo {author} {\bibfnamefont {Z.}~\bibnamefont
  {Faraei}}\ and\ \bibinfo {author} {\bibfnamefont {S.~A.}\ \bibnamefont
  {Jafari}},\ }\bibfield  {title} {\bibinfo {title} {Superconducting proximity
  in three-dimensional dirac materials: Odd-frequency, pseudoscalar,
  pseudovector, and tensor-valued superconducting orders},\ }\href
  {https://doi.org/10.1103/PhysRevB.96.134516} {\bibfield  {journal} {\bibinfo
  {journal} {Phys. Rev. B}\ }\textbf {\bibinfo {volume} {96}},\ \bibinfo
  {pages} {134516} (\bibinfo {year} {2017})}\BibitemShut {NoStop}%
\bibitem [{\citenamefont {Bergeret}\ \emph {et~al.}(2005)\citenamefont
  {Bergeret}, \citenamefont {Volkov},\ and\ \citenamefont
  {Efetov}}]{RevModPhys.77.1321}%
  \BibitemOpen
  \bibfield  {author} {\bibinfo {author} {\bibfnamefont {F.~S.}\ \bibnamefont
  {Bergeret}}, \bibinfo {author} {\bibfnamefont {A.~F.}\ \bibnamefont
  {Volkov}},\ and\ \bibinfo {author} {\bibfnamefont {K.~B.}\ \bibnamefont
  {Efetov}},\ }\bibfield  {title} {\bibinfo {title} {Odd triplet
  superconductivity and related phenomena in superconductor-ferromagnet
  structures},\ }\href {https://doi.org/10.1103/RevModPhys.77.1321} {\bibfield
  {journal} {\bibinfo  {journal} {Rev. Mod. Phys.}\ }\textbf {\bibinfo {volume}
  {77}},\ \bibinfo {pages} {1321} (\bibinfo {year} {2005})}\BibitemShut
  {NoStop}%
\bibitem [{\citenamefont {Di~Bernardo}\ \emph {et~al.}(2015)\citenamefont
  {Di~Bernardo}, \citenamefont {Salman}, \citenamefont {Wang}, \citenamefont
  {Amado}, \citenamefont {Egilmez}, \citenamefont {Flokstra}, \citenamefont
  {Suter}, \citenamefont {Lee}, \citenamefont {Zhao}, \citenamefont {Prokscha},
  \citenamefont {Morenzoni}, \citenamefont {Blamire}, \citenamefont {Linder},\
  and\ \citenamefont {Robinson}}]{PhysRevX.5.041021}%
  \BibitemOpen
  \bibfield  {author} {\bibinfo {author} {\bibfnamefont {A.}~\bibnamefont
  {Di~Bernardo}}, \bibinfo {author} {\bibfnamefont {Z.}~\bibnamefont {Salman}},
  \bibinfo {author} {\bibfnamefont {X.~L.}\ \bibnamefont {Wang}}, \bibinfo
  {author} {\bibfnamefont {M.}~\bibnamefont {Amado}}, \bibinfo {author}
  {\bibfnamefont {M.}~\bibnamefont {Egilmez}}, \bibinfo {author} {\bibfnamefont
  {M.~G.}\ \bibnamefont {Flokstra}}, \bibinfo {author} {\bibfnamefont
  {A.}~\bibnamefont {Suter}}, \bibinfo {author} {\bibfnamefont {S.~L.}\
  \bibnamefont {Lee}}, \bibinfo {author} {\bibfnamefont {J.~H.}\ \bibnamefont
  {Zhao}}, \bibinfo {author} {\bibfnamefont {T.}~\bibnamefont {Prokscha}},
  \bibinfo {author} {\bibfnamefont {E.}~\bibnamefont {Morenzoni}}, \bibinfo
  {author} {\bibfnamefont {M.~G.}\ \bibnamefont {Blamire}}, \bibinfo {author}
  {\bibfnamefont {J.}~\bibnamefont {Linder}},\ and\ \bibinfo {author}
  {\bibfnamefont {J.~W.~A.}\ \bibnamefont {Robinson}},\ }\bibfield  {title}
  {\bibinfo {title} {Intrinsic paramagnetic meissner effect due to $s$-wave
  odd-frequency superconductivity},\ }\href
  {https://doi.org/10.1103/PhysRevX.5.041021} {\bibfield  {journal} {\bibinfo
  {journal} {Phys. Rev. X}\ }\textbf {\bibinfo {volume} {5}},\ \bibinfo {pages}
  {041021} (\bibinfo {year} {2015})}\BibitemShut {NoStop}%
\bibitem [{\citenamefont {Tkachov}(2017)}]{PhysRevLett.118.016802}%
  \BibitemOpen
  \bibfield  {author} {\bibinfo {author} {\bibfnamefont {G.}~\bibnamefont
  {Tkachov}},\ }\bibfield  {title} {\bibinfo {title} {Magnetoelectric andreev
  effect due to proximity-induced nonunitary triplet superconductivity in
  helical metals},\ }\href {https://doi.org/10.1103/PhysRevLett.118.016802}
  {\bibfield  {journal} {\bibinfo  {journal} {Phys. Rev. Lett.}\ }\textbf
  {\bibinfo {volume} {118}},\ \bibinfo {pages} {016802} (\bibinfo {year}
  {2017})}\BibitemShut {NoStop}%
\end{thebibliography}
%

\end{document}


\title{Altermagnetism-induced non-collinear superconducting diode effect and unidirectional superconducting transport (Supplemental Material)}

\author{F. Yang}
\email{fzy5099@psu.edu}

\affiliation{Department of Materials Science and Engineering and Materials Research Institute, The Pennsylvania State University, University Park, PA 16802, USA}

\author{L. Q. Chen}
\email{lqc3@psu.edu}

\affiliation{Department of Materials Science and Engineering and Materials Research Institute, The Pennsylvania State University, University Park, PA 16802, USA}

\maketitle

\section{Numerical simulation}

The used specific parameters in our numerical simulations are presented in Table~\ref{TASEI}.  Notably, due to the Cooper instability of interacting fermions underlying BCS theory~\cite{abrikosov2012methods,bardeen1957theory,schrieffer1964theory,mahan2013many}, the momentum summation over pairing electrons is restricted to states near the Fermi surface. Specifically, the spin-up electron with momentum ${\bf k+q}$ and the spin-down electron with momentum ${\bf -k+q}$ are both confined within their respective Debye shells, as determined by the spin-dependent condition $|\xi_{s{\bf k+q}}+sh_{s{\bf k+q}}|\le\omega_D$. In BCS theory, the Cooper instability arises only for fermions near the Fermi surface, where a weak attractive interaction, typically phonon-mediated, is effective. This constraint reflects the limited range of the fermionic  pairing interaction and imposes an energy cutoff that restricts pairing to low-energy electrons~\cite{bardeen1957theory}, prevents ultraviolet divergence in the gap equation~\cite{abrikosov2012methods,mahan2013many}, and helps preserve approximate particle–hole symmetry of the condensate~\cite{pekker2015amplitude,yang19gauge,yang2018gauge}. However, previous microscopic treatments~\cite{PhysRevB.111.054501,hu2025unconventional} have overlooked this crucial aspect and performed the pairing summation over the entire Brillouin zone. This practice neglects the physical energy cutoff imposed by the pairing mechanism and may artificially overestimate the available phase space, particularly in systems with spin-split bands or strong momentum dependence. One consequence of this oversight is that, while the BCS–FF transition is continuous in both our work and Refs.~\cite{PhysRevB.111.054501,hu2025unconventional}, the FF–normal transition is also reported to be continuous in Refs.~\cite{PhysRevB.111.054501,hu2025unconventional}, whereas our results show a clear first-order character, reflecting the abrupt/sharp collapse of the condensate at large finite pairing momentum as reported in Ref.~\cite{PhysRevB.111.174507}. Moreover, the superconducting gap considered in both Refs.~\cite{PhysRevB.111.054501,hu2025unconventional} appears comparable to the Fermi energy, and thus, is unrealistically large for conventional phonon-mediated superconductors. 

In realistic heterostructures, the strength of the altermagnetic proximity effect can vary significantly due to interface quality, layer thickness, and growth conditions. Therefore, in our study, we treat the altermagnetic spin splitting strength as a tunable parameter. First-principles calculations have confirmed that intrinsic altermagnetic spin splittings in bulk materials can reach values on the order of hundreds of meV~\cite{vsmejkal2022emerging,noda2016momentum}. However, in proximity systems, the induced spin splitting in the adjacent superconducting layer is expected to be substantially smaller, owing to the short-range nature of exchange coupling and the presence of interface mismatch, disorder, or orbital filtering. The scale of such proximity-induced spin splitting is often comparable to or smaller than the superconducting gap itself~\cite{PhysRevB.95.075304,PhysRevB.84.144522,PhysRevB.91.165425,PhysRevLett.118.117001,RevModPhys.77.935}, which typically lies in the meV range. Accordingly, the range of the splitting values we consider (0–2~meV) in the main text should not only be able to capture realistic interfacial effects but also allow us to explore the crossover from weak to moderately strong proximity coupling in a physically relevant regime.

\begin{center}
\begin{table}[htb]
  \caption{Specific parameters used in the numerical simulations.
    The pairing interaction strength $U$ is determined by requiring the zero-temperature superconducting gap $\Delta_0$ to match a prescribed value of SC thin film NbN~\cite{chockalingam2008superconducting} in the absence of spin splitting. The Debye cutoff $\omega_D$ is chosen to yield a critical temperature $T_c = 13$~K~\cite{yang24thermodynamic,chockalingam2008superconducting,matsunaga2014light} under the same conditions. The effective mass $m$ is set equal to the free electron mass $m_e$. Moreover, in conventional superconductors, the gap magnitude satisfies $\Delta_0 \ll E_F$. In this regime, the results presented in the normalized form in the main text are independent of the specific choice of $E_F$. }  
\renewcommand\arraystretch{1.5}   \label{TASEI}
\begin{tabular}{l l l l l l l l}
    \hline
    \hline
         model parameters &\quad\quad\quad~$\Delta_0$&\quad\quad\quad~$\omega_D$&\quad\quad~$m/m_e$\\  
    &\quad\quad\quad2~meV&\quad\quad$0.01~$meV&\quad\quad$0.998$\\   
    \hline
    \hline
\end{tabular}\\
\end{table}
\end{center}

\begin{figure}[htb]
  {\includegraphics[width=8.7cm]{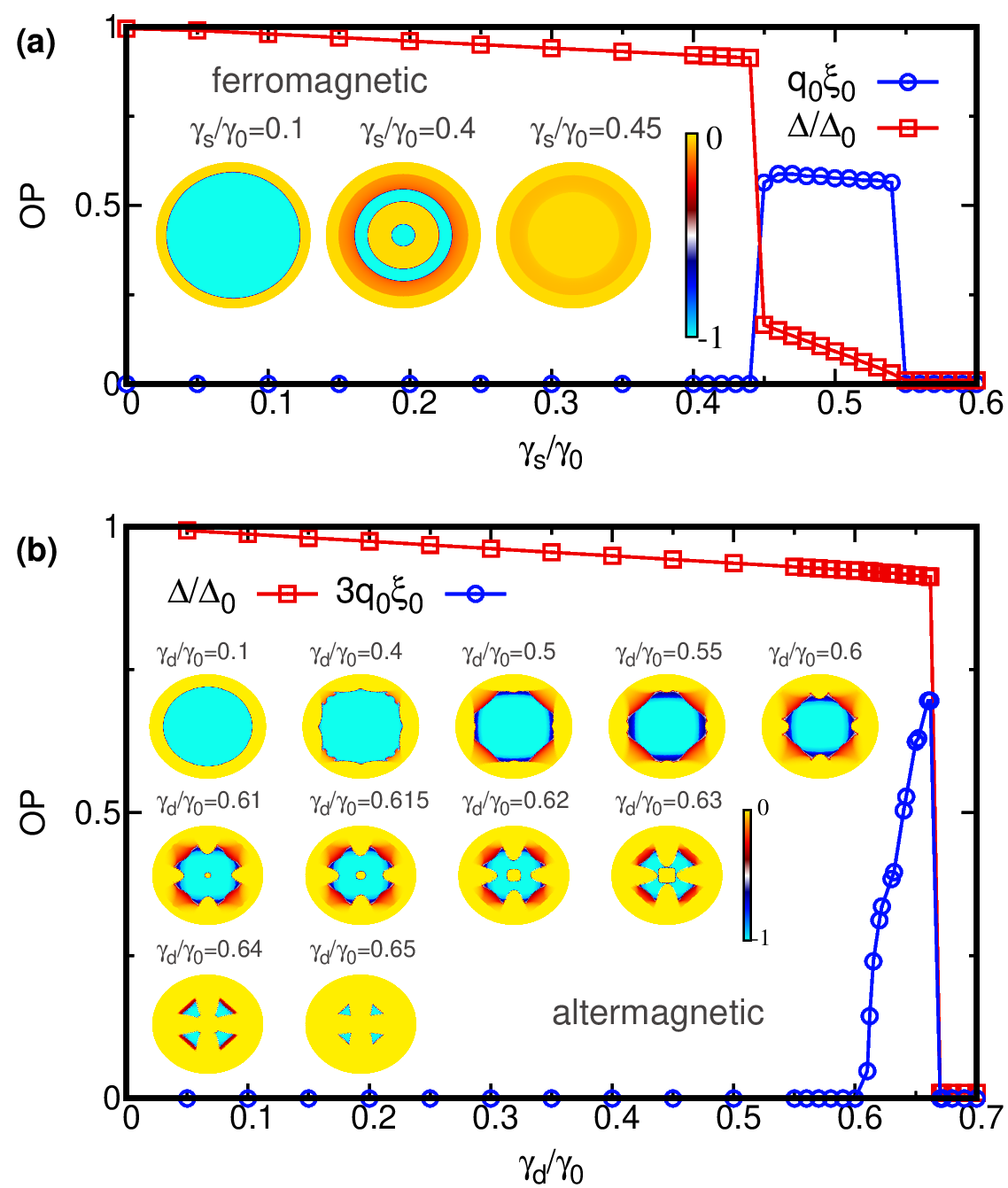}}
  \caption{Simulated order-parameter (OP) results for the optimal CM momentum $q_0$ and the corresponding practical SC gap $\Delta = \Delta_{\bf q=q_0}$ as functions of the spin-splitting strength for {\bf (a)} ferromagnetic and {\bf (b)} $d$-wave altermagnetic proximity, same as Fig.~2 in the main text. Insets: the free-energy landscape plotted in $q_x$–$q_y$ plane. The plotted $q$ range extends up to $0.8q_0$ in each direction. $T=0.1~$K. Normalized parameters used here are: the zero-splitting gap $\Delta_0=2~$meV, coherence length $\xi_0 = v_F/\Delta_0$, and splitting unit $\gamma_0 = \Delta_0/k_F^2$.} 
\label{Sfigyc1}
\end{figure}

We plot the free-energy landscape in the insets of Fig.~\ref{Sfigyc1} in comparison with the insets of Fig.~2 in the main text, to show that the CM momentum dependence of $\Delta_{\bf q}$ directly reflects the inverse of the normalized free-energy landscape as the condensation energy satisfies $\mathcal{F}^s_{\mathbf{q}} - \mathcal{F}^n_{\mathbf{q}} \propto -\Delta_{\bf q}^2$.  

\section{Spin-triplet state}

Finally, we discuss the possible emergence of spin-triplet pairing correlations. In the presence of a generalized momentum-dependent spin-splitting vector field ${\bf \Omega}_{\bf k}$ and a finite center-of-mass (CM) momentum ${\bf q}$ of the Cooper pairs, we introduce the Nambu spinor {\small {${\hat \Phi}_{\bf
  k}=[\phi_{\uparrow{\bf k+q}},\phi_{\downarrow{\bf 
    k+q}},\phi^{\dagger}_{\uparrow{\bf -k+q}},\phi^{\dagger}_{\downarrow{\bf
  -k+q}}]^{T}$}}, and write the BdG Hamiltonian for an $s$-wave superconductor as~\cite{abrikosov2012methods} 
\begin{equation}
\label{Hamiltonian}
{\hat H_{S}}=\frac{1}{2}\int{d{\bf k}}{\hat \Phi}^{\dagger}_{\bf
  k}{\hat H_s}({\bf k})\rho_3{\hat \Phi}_{\bf k},
\end{equation}
with
\begin{eqnarray}
&&{\hat H_{s}}({\bf k})=\left(\begin{array}{cc}
\xi_{\bf k^+}+{\bf \Omega}_{\bf k^+}\cdot{\bm \sigma} & \Delta_{\bf q}i\sigma_2\\
\Delta^*_{\bf q}i\sigma_2 & \xi_{\bf k^-}+{\bf \Omega}_{\bf k^-}\cdot{\bm
\sigma}\\
\end{array}\right).~~~~~
\end{eqnarray}
Here, $\rho_3=\sigma_0\otimes\tau_3$; $\sigma_i$ and $\tau_i$ stand for the Pauli
matrices in spin and particle-hole spaces, respectively; ${\bf k}^{\pm}=\pm{\bf
  k+q}$ with ${\bf q}$ standing for the CM momentum.  The equilibrium Green’s function in Nambu $\otimes$ spin space is defined as~\cite{abrikosov2012methods} 
\begin{equation}
G_{\bf q}({\bf k},{\tau})=-\rho_3\langle{T_{\tau}{\hat \Phi}_{\bf
  k}(\tau){\hat \Phi}^{\dagger}_{\bf
  k}(0)}\rangle, 
\end{equation}
where $T_{\tau}$ represents the chronological-ordering operator and $\tau$ is
the imaginary time. By expressing 
\begin{equation}
G_{\bf q}({\bf k},\tau)=\left(\begin{array}{cc}
g_{\bf q}({\bf k},\tau)& f_{\bf q}({\bf k},\tau)\\ 
f^{\dagger}_{\bf q}({\bf -k},\tau)& {g^{\dagger}}_{\bf q}({\bf -k},\tau)\\ 
\end{array}\right),
\end{equation} 
one can obtain the normal Green function $g_{\bf q}({\bf k},\tau)$ and anomalous Green
function $f_{\bf q}({\bf k},\tau)$.

In the Matsubara
representation $G_{\bf
  k}(i\omega_n)=\int^{\beta}_{0}d{\tau}e^{i\omega_n{\tau}}G_{\bf k}(\tau)$, the Gorkov equations  $[i\omega_n\rho_3-H_{s}({\bf k})]G_{\bf q}({\bf k},i\omega_n)=1$ yield:
\begin{equation}
  \label{pro1}
(i\omega_n-\xi_{\bf k^+}-{\bf \Omega}_{\bf k^+}\cdot{\bm \sigma})f_{\bf q}({\bf
    k},i\omega_n)-\Delta_{\bf q}i\sigma_2g^{\dagger}_{\bf q}({\bf
    k},i\omega_n)=0,
\end{equation}
\begin{equation}
  \label{pro2}
-\Delta^*_{\bf q}i\sigma_2f_{\bf q}({\bf
    k},i\omega_n)+(-i\omega_n-\xi_{\bf k^-}-{\bf \Omega}_{\bf k^-}\cdot{\bm
\sigma})g^{\dagger}_{\bf q}({\bf
    k},i\omega_n)=1.  
\end{equation}
Here, $\omega_n=(2n+1)\pi{k_BT}$ are the Matsubara frequencies. Multiplying
Eq.~(\ref{pro2}) by $i\sigma_2$ from the left side, one
immediately has
\begin{equation}
  \label{pro3}
\Delta_{\bf q}^*f_{\bf q}({\bf
    k},i\omega_n)+(-i\omega_n-\xi_{\bf k^-}+{\bf \Omega}_{\bf k^-}\cdot{\bm
\sigma}^*)i\sigma_2g^{\dagger}_{\bf q}({\bf
    k},i\omega_n)=i\sigma_2.
\end{equation}
Then, using Eq.~(\ref{pro1}) to replace $i\sigma_2g^{\dagger}_{\bf q}({\bf k},i\omega_n)$ in Eq.~(\ref{pro3}), the
anomalous Green function can be obtained:  
\begin{equation}
  \label{pairing}
f_{\bf q}({\bf k},i\omega_n)=[f^s_{\bf q}({\bf k},i\omega_n)+{\bf f}^t_{\bf
  q}({\bf k},i\omega_n)\cdot{\bm \sigma}]i\sigma_2,
\end{equation}
with the singlet $f^s_{\bf q}({\bf k},i\omega_n)$ and triplet ${\bf f}^t_{\bf
  q}({\bf
  k},i\omega_n)=\big(\frac{f_{\downarrow\downarrow}-f_{\uparrow\uparrow}}{2},\frac{f_{\downarrow\downarrow}+f_{\uparrow\uparrow}}{2i},f_{\downarrow\uparrow+\uparrow\downarrow}\big)$
pairings~\cite{RevModPhys.77.935,PhysRevB.95.075304} 
written as:  
\begin{eqnarray}
f^s_{\bf q}({\bf k},i\omega_n)&=&\Delta_{\bf q}\frac{|\Delta_{\bf q}|^2-(i\omega_n-\xi_{\bf k^+})(i\omega_n+\xi_{\bf
  k^-})-{\bm \Omega}_{\bf k^+}\cdot{\bm \Omega}_{\bf k^-}}{{\prod_{\mu=\pm}(i\omega_n-E^e_{\mu{\bf k}})(i\omega_n-E^h_{\mu{\bf k}})}},\\
{\bf f}^t_{\bf q}({\bf k},i\omega_n)&=&\Delta_{\bf q}\frac{i{\bm \Omega}_{\bf k^+}\times{\bm \Omega}_{\bf k^-}-(i\omega_n-\xi_{\bf k^+}){\bm \Omega}_{\bf 
  k^-}-(i\omega_n+\xi_{\bf 
  k^-}){\bm \Omega}_{\bf k^+}}{{\prod_{\mu=\pm}(i\omega_n-E^e_{\mu{\bf k}})(i\omega_n-E^h_{\mu{\bf k}})}},\label{tp}
\end{eqnarray}
where $E^{e(h)}_{\mu{\bf k}}$ ($\mu=\pm$) stand for the quasiparticle electron (hole)
energy spectra in superconductors.

Clearly, Eq.~(\ref{tp}) shows that spin-triplet correlation/pairing emerges naturally due to the presence of momentum-dependent spin splitting. For a $d$-wave altermagnetic spin splitting, i.e., ${\bf \Omega}_{\bf k}=h_{\bf k}{\bf {\hat z}}$, the BCS state with ${\bf q} = 0$ hosts the odd-frequency, even-parity ($d$-wave) triplet pairing ${\bf f}^t_{\bf q=0}({\bf k},\omega)\propto\Delta_{\bf q}{\omega}h_{\bf k}{\bf {\hat z}}$. Upon entering the FF state (${\bf q} \ne 0$), additional component  ${\bf k}\cdot{\bf q}h_{\bf k}{\bf {\hat z}}\Delta_{\bf q}$ in ${\bf f}^t_{\bf q}({\bf k},\omega)$ emerge, giving rise to the even-frequency, odd-parity spin-triplet correlations. The emerging  triplet components in superconducting anomalous correlation may lead to novel phenomena such as an anomalous paramagnetic  Meissner effect, as introduced in Refs.~\cite{RevModPhys.77.1321,RevModPhys.77.935} and experimentally observed in Ref.~\cite{PhysRevX.5.041021}, and cause the magnetoelectric effect~\cite{yang2018fulde,PhysRevLett.118.016802}. All of these features here, with/without the CM momentum, agree fully with the symmetry analysis presented in Ref.~\cite{PhysRevB.95.075304}.

However, it is important to emphasize that the presence of spin-triplet pairing correlations does not imply the emergence of a spin-triplet order parameter. {Importantly, these induced triplet channels do not affect any of the main results of the present work, which focuses on the superconducting phase evolution of the order parameter as a function of temperature and altermagnetic splitting, as well as on the phase transition and its critical behavior (including critical currents). The reason is that in conventional superconducting metals, the pairing interaction is typically dominated by $s$-wave phonon-mediated attraction and favors a spin-singlet channel. Thus, while proximity-induced spin splitting can admix triplet components into the anomalous propagator, the superconducting order parameter itself remains in the singlet sector. More specifically,  the gap equation employed in our analysis,
\begin{equation}
{\hat \Delta}_{\bf q}=U\sum_{{\bf k},\omega}f_{\bf q}({\bf k},\omega)
\end{equation}
is based on a momentum-independent, spin-singlet pairing potential $U$ (conventional superconducting metals). Such a $s$-wave pairing interaction energetically selects the singlet, even-frequency channel, and projects out odd-frequency or odd-parity components from contributing to the self-consistent gap amplitude, yielding an $s$-wave  spin-singlet gap
\begin{equation}
{\hat \Delta}_{\bf q}=\Delta_{\bf q}i\sigma_2,~~~~~\Delta_{\bf q}=U\sum_{{\bf k},\omega}f^s_{\bf q}({\bf k},\omega). 
\end{equation}
  As a result, while symmetry-allowed triplet correlations may appear in the anomalous Green’s function, they do not feed back into the singlet gap equation and therefore do not modify the superconducting phase diagram, the BCS–FFLO transition, or the resulting critical currents analyzed in this work. For comparison, we also note that in one of our earlier works~\cite{PhysRevB.95.075304}, where we studied a special layered system with a fully frequency- and momentum-dependent pairing interaction $U_{\mathbf{k}\mathbf{k}'}(\omega,\omega')$, the gap equation took the form ${\hat \Delta}(\mathbf{k},\omega)=\sum_{\mathbf{k}',\omega'} U_{\mathbf{k}\mathbf{k}'}(\omega,\omega') f(\mathbf{k}',\omega')$. In that case, all pairing channels (singlet, triplet, even- and odd-frequency) enter the self-consistent equation and can influence the critical behavior.}

%